\documentclass[aps,prfluids,titlepage,showpacs,a4paper,preprint,longbibliography,
tightenlines,superscriptaddress,nofootinbib]{revtex4-2}
\usepackage{graphicx,xcolor}
\usepackage{amsmath}

\usepackage{epsfig,latexsym,array,amssymb}
\newcommand{\be}{\begin{equation}}
\newcommand{\ee}{\end{equation}}
\newcommand{\ba}{\begin{eqnarray}}
\newcommand{\ea}{\end{eqnarray}}
\newcommand{\benn}{\begin{displaymath}}
\newcommand{\eenn}{\end{displaymath}}
\renewcommand{\d}[1]{{\rm d}#1}

\renewcommand{\vec}[1]{\mbox{\boldmath $#1$}}

\newcommand{\svec}[1]{\mbox{\scriptsize \boldmath ${#1}$}}
\newcommand{\mat}[1]{\mbox{\boldmath ${\rm #1}$}}

\newcommand{\wt}[1]{\widetilde{#1}}

\reversemarginpar
\newcommand{\nonote}[1]{ } % nomargin note
\newcommand{\deleted}[1]{\color{red} #1 \color{black}}
\renewcommand{\deleted}[1]{}
\newcommand{\tnti}{turbulent-nonturbulent interface}
\newcommand{\tti}{turbulent-turbulent interface}

\begin{document}
%
%\title{The turbulent/non-turbulent interface:\\a coherent structure ?}
%\title{Relating turbulent interfaces to coherent structures}
\title{Coherent Structures Governing Transport 
%%% through 
at
Turbulent
Interfaces}
\author{Ali R Khojasteh, Lyke E. van Dalen, Coen Been, Jerry
Westerweel and Willem van de Water}

\affiliation{
Laboratory for Aero- and Hydrodynamics, Delft University of Technology
and J.M. Burgers Centre for Fluid Dynamics, 2628 CD Delft, The
Netherlands}
\email[Corresponding author: ]{A.R.Khojasteh@tudelft.nl}
\date{12 December 2024}
\begin{abstract}
In an experiment on a turbulent jet, we detect interfacial turbulent layers in a frame that moves, on average, along with the  \tnti.  This significantly prolongs the observation time of scalar and velocity structures and enables the measurement of two types of Lagrangian coherent structures.  One structure, the finite-time Lyapunov field (FTLE), quantifies advective transport barriers of fluid parcels while the other structure highlights barriers of diffusive momentum transport.  These two complementary structures depend on large-scale and small-scale motion and are therefore associated with the growth of the turbulent region through engulfment or nibbling, respectively.
We detect the \tnti\ from cluster analysis, where we divide the measured scalar field into four clusters.  Not only the \tnti\ can be found this way, but also the next, internal, turbulent-turbulent interface.
Conditional averages show that these interfaces are correlated with barriers of advective and diffusive transport when the Lagrangian integration time is smaller than the integral time scale. Diffusive structures decorrelate faster since they have a smaller timescale. 
Conditional averages of these structures at internal turbulent-turbulent interfaces show the same pattern with a more pronounced jump at the interface indicative of a shear layer. This is quite an unexpected outcome, as the internal interface is now defined not by the presence or absence of vorticity, but by conditional vorticity corresponding to two uniform concentration zones.
% The diffusive normal flux at the interface agrees with the idea of so-called `nibbling' where the turbulent domain propagates outward through viscous diffusion transport of vorticity.
The long-time diffusive momentum flux along Lagrangian paths represents the growth of the turbulent flow into the irrotational domain, a direct demonstration of nibbling.  
The diffusive flux parallel to the \tnti\ appears to be concentrated in a diffusive superlayer whose width is comparable with the Taylor microscale, which is relatively invariant in time. 
%

% \marginpar{The abstract should also mention the result on \tti.}
%
\end{abstract}
\maketitle

\section{Introduction}
\begin{figure}[ht]
\centering
\includegraphics[scale = 0.9]{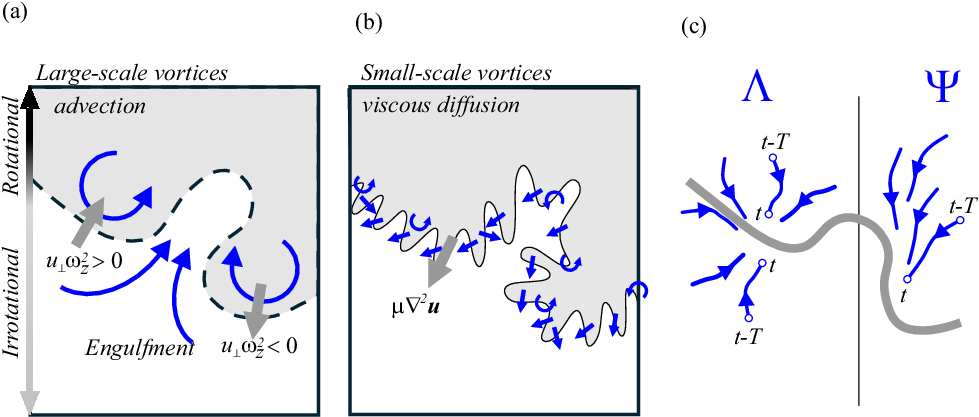}
\caption{
(a) The \tnti\ separates irrotational and turbulent flow.  A
large-scale process, i.e. engulfment, mixes irrotational fluid into the
turbulent domain.
(b) Small-scale vortices propagate and convolute the \tnti, enabling 
%%% .  After the wrinkles caused by them, there is 
a diffusive momentum flux $\mu \nabla^2
\vec{u}$ across the interface.
(c) Advection-diffusion at the interface in a Lagrangian frame. The
two distinct types of structures considered in this paper are: the
finite-time Lyapunov field $\Lambda$, i.e. (i) the divergence (in
backward time) of fluid parcels that are close at time $t$, and (ii) the
convergence $\Psi$ of streamlines of the diffusive momentum flux.
Both quantities are averages along Lagrangian trajectories over a
time $T$ preceding the instant of observation. }
\label{fig.cartoon}
\end{figure}

%....................................................................
A turbulent flow can be viewed as regions of uniform momentum separated by interfacial layers where the gradient of vorticity fluctuates strongest \citep{Corssin1955,Meinhart1995,Adrian2000,Ishihara2013,Eisma2015pof}. In jet flow, the outermost of these layers is known as the \tnti, which separates rotational (turbulent) and irrotational regions, and is characterized by a sharp change in flow properties \citep{Westerweel2005,Silva2014}.  At the interface, non-turbulent fluid is incorporated into the turbulent region, by both large scale and small scale processes, referred to as `engulfment' and `nibbling' respectively. Numerical and experimental findings indicated that 
%%% \cite{Westerweel2005,Westerweel2009,Mathew2002} 
%%% concluded that 
the entrainment process is predominantly a small-scale process, with engulfment contributing only slightly in
the self-similar region of the jet \cite{Mathew2002,Westerweel2005,Westerweel2009,Silva2014}.
%
% A cartoon in 
Figure~\ref{fig.cartoon} illustrates the current state of
affairs, and sketches the focus of the present article.  Engulfment
involves fluid motion on large scales, while small-scale vortices,
concentrated in a vortical superlayer \citep{Corssin1955}, dominate
the spread of the turbulence into the irrotational domain. In Figure~\
\ref{fig.cartoon}(b), the flow of enstrophy stops at the \tnti, but
the small-scale vortices propel the flow of viscous momentum $\mu
\nabla^2 \vec{u}$. 
While engulfment and nibbling have so far been studied in the
Eulerian frame, we emphasize their Lagrangian context. It 
%%% has
inspired the design of our quasi-Lagrangian setup where the detection
of velocity and scalar fields moves with the average interface
velocity. 
Figure~\ref{fig.cartoon}(c) illustrates two (complementary)
quantities of interest in this paper: the backward-in-time 
rate of separation ($\Lambda$) of
two fluid parcels, 
%%% ($\Lambda$), 
and the convergence ($\Psi$) of
viscous momentum flux $\mu \nabla^2 \vec{u}$.  These Lagrangian
structures are typically defined for a finite time $T$.  In this paper
these times {\em precede} the instant of observation.
%

%--------------------------------------------------------------------
%--------------------------------------------------------------------
% \marginpar{Somewhere in the Results section you should show these streamlines of the diffusive momentum flux!}

The state of the art in recent experiments on turbulent interfaces
involves the simultaneous measurement of the velocity field using
particle-image velocimetry (PIV), while the concentration field is
measured using laser-induced fluorescence (LIF)
\citep{Mistry2016,Mistry2019,Buxton2023}. It enables the detection
of the \tnti\ of a jet seeded with dye and 
%%% find 
determine its correlation with
the velocity field. However, flows at high Reynolds numbers pose a challenge
for achieving high spatial resolution, restricting the detail and
accuracy of the observations.  Furthermore, in most experiments the flows are
recorded with stationary cameras, which limits the possibility to track Lagrangian evolution of these interfaces 
%%% dynamics
over longer times.

In this work we consider the \tnti\ of a submerged turbulent round jet exiting
into a quiescent volume of 
%%% still 
fluid at a Reynolds number ${\rm Re} \approx
1.25\times10^4$.  The jet is seeded with dye, which is visualized
using LIF, while the velocity field in an axial planar cross section is measured
using PIV.  A novelty of this experiment is the measurement of the
dye concentration and velocity in a frame that moves with the average downstream and radial velocity of the 
\tnti.
%%% \ velocity. 
Using the statistics of the measured
concentration fields, we identify not only the \tnti\ but also an
{\em internal} \tti. 
% interface: the \tti.  

%%% Our 
The purpose of this investigation 
%%%
is to relate these interfaces to coherent Lagrangian
structures.  One particular structure, i.e. the ridge-like local maxima of the
finite-time Lyapunov field $\Lambda(\vec{x})$, forms a barrier to
large-scale advective transport
\citep{Haller2000,Shadden2005,Haller2015}. The other structure,
$\Psi(\vec{x})$, is related to barriers of the diffusive flux of
momentum, and thus highlights small-scale structures
\citep{Haller2020}.  Both 
%%% kinds 
structures
%%%
are {\em objective}: they are
independent of the frame of observation \citep{Haller2020}.
The detection of these structures requires 
%%% long 
extended 
%%%
observation times in
a Lagrangian frame, which necessitates an experimental setup where
the detection 
%%% goes with
cameras move along 
%%%
the flow.
%
%%% By moving the field of view along with the interface at its averagevelocity, we enhance 
This provides a substantial enhancement of
%%%
the spatial resolution of the 
%%%
measured 
%%%
velocity field
%%% reconstruction 
near the interface over 
%%% a large 
an extended 
%%%
downstream distance.
This method 
%%% surpasses 
avoids limitations in 
%%% the 
spatial resolution 
%%% limitations of 
associated with 
%%%
a fixed
camera observing the entire jet, as well as scenarios where the
camera measures near the interface at a fixed location, which fail to
capture the long-time evolution of the interface. The present paper
(and our earlier work, \cite{Jesse2021}) focuses on 
%%%
the
%%%
time dependence,
whereas existing studies involve snapshots only.

The prevalence of small-scale nibbling over large-scale engulfment
was concluded on the basis of a small correlation length (of the
order of the Taylor microscale) of the velocity fluctuations near the
\tnti\ and the relatively small area of irrotational fluid inside a
planar cross secion of the jet
\citep{Westerweel2005,Westerweel2009,Silva2014}. 
%%%
As noted by \citet{Mathew2002}, the entrainment process is related across scales.  
%%%
\citet{Mistry2016}
formulate a corollary to these results: the \tnti\ is 
%%%
considered to be 
%%%
a fractal
surface, and using filtering with an increasing filter length
$\Delta$ they conclude that the filtered entrainment velocity
increases with increasing $\Delta$, while the filtered surface area
decreases with increasing $\Delta$, such that the mass flux (the
product of entrainment velocity and area) does not depend on
$\Delta$, which was already suggested by \citet{Meneveau1990}.

% In this sense, neither nibbling nor engulfment prevails.  
% (not liked by Jerry)
%%% then why mention it????
%%% Mathew & Basu already mention that both large-scale and small-scale motion in turbulence are evidently linked, so that a process of entrainments at small scales is linked to a process at large scales
%%% I think we can mention it but clearly in this type of context
%%% I added one sentence ...

% Vortical Lagrangian coherent structures -- averages of the vorticity
% field along Lagrangian paths have been related to the \tnti\ of a
% gravity current by \cite{Neamtu2019}.  Conditional averages
% demonstrated their effect on the \tnti\ and entrainment dynamics.  

The \tnti\ is the boundary between the turbulent (rotational) domain
and the nonturbulent (irrotational) domain, which we will sometimes
denote as `blue sky' below (as in a white turbulent cloud against a blue sky background). 
%%% in the sequel.
%%% 'It' should refer to the tnti, not the 'blue sky'
%%% It 
The \tnti\ is not a material surface
and propagates into the irrotational domain with velocity $E_B = -2
V$ for a round turbulent jet \citep{Turner1986}, where $V$ is the velocity of a fluid parcel perpendicular to the
interface.
The role of viscous and nonviscous effects can be appreciated by
considering the propagation velocity $E_B$ of the \tnti.
%%% This was not invented by me!! Use the original refs!!!
\citet{Westerweel2005,Westerweel2009} argue that 
the entrainment boundary velocity 
$E_B$ follows from a
nonviscous stress balance, $E_B \Delta \wt{U} \approx -\left\langle
\wt{u}\wt{v} \right\rangle$ (the boundary jump condition), with
$\wt{u}, \wt{v}$ fluctuating velocities 
%%% in the frame of the
conditional to the location of the interface, and $\Delta \wt{U}$ the jump of the mean axial velocity
component 
\citep{Kovasznay1967,Kovasznay1970,Reynolds1972,Westerweel2009}.
In this frame the enstrophy does not change and an analysis of the
enstrophy transport equation \citep{Holzner2011,Watanabe2014} also
allows to isolate the viscous contributions to the interface
velocity.
After all, the initial transport of vorticity away from a body
accelerated from rest is through viscous diffusion
\citep{Batchelor1967}.

%%% repeat
% The \tnti\ is a discontinuity, expressed in the boundary jump condition \citep{Westerweel2005,Westerweel2009}.
%
% \note{Ali, I hope this is what you had in mind.}

% Recently, flow structures have been identified that characterize
% transport of scalar and momentum in a turbulent flow.  These
% structures are defined in the Lagrangian frame, and require
% experiments in which the detection also moves with the flow.  

% These two complementary effects are embodied in the two
% complementary structures on which we will focus our experiments. 

While the \tnti\ separates turbulent from nonturbulent fluid, {\em
internal} interfaces were identified by \citet{Eisma2015pof} using
thresholds on a velocity gradient tensor \citep{Kolar2007}.  These
thin shear layers were first reported 
%%% described 
by \citet{Meinhart1995}, and 
%%% were
subsequently studied in turbulent boundary layers 
\citep{Adrian2000,Chauhan2014,Asadi2022}.
%%% (which contains extensive references).  
In all these cases,
interfaces were found using measured velocity fields, either by
imposing thresholds, or by identifying zones of approximately uniform
momentum using PDF's of the instantaneous streamwise velocity.  

In the present article we 
%%% find 
identify
%%%
interfaces from the scalar field using
the well-established techniques of cluster analysis
\citep{Bezdek1980,Fan2019}.  This is obvious for the \tnti, which
separates scalar from the absence of scalar. However, the scalar
concentration field 
%%% inside 
towards
%%%
the core of the jet 
%%% is 
appears to be 
%%%
organized in
uniform concentration zones 
%%% which 
that
%%%
can be used to find {\em internal}
boundaries.  Specifically, when these zones are ranked according to
their concentration level, the \tnti\ is the boundary between the
first two zones, with the first zone the region of unmixed fluid,
whereas a \tti\ is a boundary between subsequent zones.  We then
compute conditional averages of vorticity $\omega$, and the new
fields $\Lambda, \Psi$ on these internal boundaries.

%%%
An outline of this paper is as follows:
%%%
Coherent structures are described in Sec.\ \ref{sec:structures}. The
experimental setup, where we move with the flow, is discussed in
Sec.~\ref{sec:setup}; it includes a discussion (Sec.\
\ref{sec:move}) of how to interpret results in a moving frame.    
The detection of the turbulent interfacial layers based on images of dye fluorescence and
cluster analysis is 
%%%
described
%%%
in Sec.~\ref{sec:find.tnti}. Conditional
averages on the contorted fractal interfaces are defined in Sec.~\ref{sec:cond}. The results are presented 
% given 
in Sec.~\ref{sec:results}, and finally the conclusions of this work are given in Sec.~\ref{sec:conclusions}.

%....................................................... Theory mania
%....................................................................
\section{Coherent structures}
\label{sec:structures}
We view coherent structures in this work as barriers for either
advective or diffusive quantities: manifolds that hinder momentum
transport. In the case of diffusive transport, it is possible to quantify
the flux across an interface, time averaged along Lagrangian paths,
using conditional averages. 
%

%%%
% \marginpar{Move this to Sec.II}
The significance of the two Lagrangian structures introduced in figure~\ref{fig.cartoon}c with respect to the \tnti\ is as follows:
%%%
% \marginpar{JW: too simplistic??}
if the transport across the \tnti\ is carried by large-scale
structures, the interface should not be related to barriers of
momentum flux. On the other hand, if the growth of the turbulent
region is through diffusion of vorticity, the \tnti\ should not be a
barrier to diffusive momentum flux.  
The possible association with a barrier field is but one aspect of
the \tnti.  We now briefly describe the two coherent structures that are measured
in our experiments. 

\subsection{Barriers for advective transport}
\label{sec:ftle}
Finite-time Lyapunov exponents gauge the exponentially fast spreading
of nearby fluid parcels.  Ridge-like maxima 
%%%
in the finite-time Lyapunov exponent field $\Lambda(\vec{x}, t)$
%%% -large-scale features -
of the associated field $\Lambda(\vec{x}, t)$ 
form barriers for passive tracers
\citep{Shadden2005,Haller2015}, 
which are associated with the large-scale structure of the flow.
%%%

The evolution operator (flow map) $\mat{F}$ of material points
$\vec{x}(t)$ that start at $\vec{x}_0$ and are carried by the
velocity field $\vec{u}(\vec{x}, t)$ is defined as
\be
   \vec{x}(t) = \mat{F}_{t_0}^t (\vec{x}_0) = \int_{t_0}^t
   \vec{u}(\vec{x}(t'), t') \: \d t'.
\ee
Its gradient field $\mat{M}_{t_0}^t = \nabla \mat{F}_{t_0}^t$
describes the evolution of small separations $\vec{\delta}$ between
fluid parcels.  It can be computed from a measured velocity field by
integrating the evolution of a vector $\vec{\delta}$ in the velocity
gradient field along a Lagrangian trajectory, 
\be
   \frac{\d \vec{\delta}}{\d t} = \mat{A}(\vec{x}(t), t)
   \cdot \vec{\delta}(t), 
   \;\; \mbox{with} \;\;
   \frac{\d \vec{x}(t)}{\d t} = \vec{u}(\vec{x}, t), 
   \mat{A} = \nabla \vec{u} \;\;\mbox{and} \;\;
   \vec{x}(t = t_0) = \vec{x}_0.
\label{eq:lcs}
\ee
The largest eigenvalue $\lambda_2$ of the positive Cauchy-Green tensor,
\be
   \mat{C}_{t_0}^t = \mat{M}_{t_0}^t \: 
   \left(\mat{M}_{t_0}^t\right)^\dagger,
\label{eq:cgt}
\ee
with $t = t_0 + T$ and with $\dagger$ the adjoint operation, then
defines the finite-time Lyapunov field $\Lambda_T(\vec{x}_0, t_0)$ as
\be
   \Lambda_{T}(\vec{x}_0, t_0) = \frac{1}{2 |T|} \ln(\lambda_2).
\label{eq:lya}
\ee   
Our experimental technique gives access to the planar cross section of
the velocity field; consequently, there are only two eigenvalues.
In the case $T > 0$, Eq.\ (\ref{eq:cgt}) expresses the separation of
fluid parcels that are close at $t_0$ and separate in the future $t_0
+ T$.  Similarly, by integrating the trajectories {\em backward} in
time, the largest eigenvalue of $\mat{C}_{t_0}^{t_0-T}$ defines the
{\em backward} Lyapunov field $\Lambda_{-T}(\vec{x}_0, t_0)$.
\citet{Jesse2021} found that it was the backward in time field
$\Lambda_{-T}$ that delineated large-scale structure of the scalar
field in the core of 
%%% the 
a
%%%
jet. 
We expect that in the irrotational domain the separation 
%%% will be
remains
%%%
small, while there 
%%% will be 
a sudden increase 
%%%
occurs
%%%%
when a fluid parcel 
%%% will enter 
enters the turbulent flow region.
Since the field $\Lambda_{-T}(\vec{x}, t)$ is Lagrangian, it is {\em
objective}: it is the same for all observers, independent of their
(moving, accelerated) observation frame.
% It therefore is a worthy candidate coherent structure.
%
%%% \nonote{Lyke: this indeed was a ``first time'': the relation between uniform concentration zones and ftle.}
%
To emphasize its ridges, the field $\Lambda_{-T}$ 
%%% has been 
is 
%%%
filtered
to include only regions with negative curvature in the direction of
the eigenvector corresponding to $\lambda_2$.

%--------------------------------------------------------------------
\begin{figure}[ht]
\centering
\includegraphics[scale = 0.9]{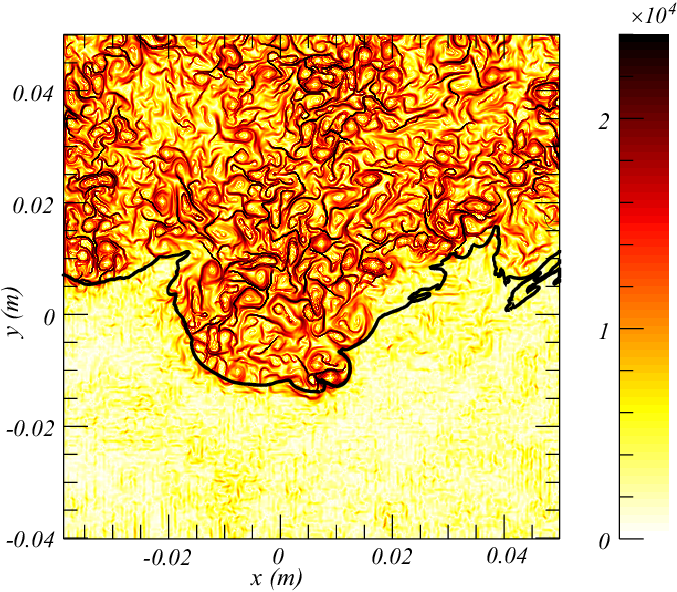}
\caption{
Snapshot of the diffusive flux field $\Psi_0$; the solid line represents the \tnti\ as defined by the fluorescent dye.
% zoomed in
%
% gray line: interface time-shifted by $\Delta t = 0.22 L
% / U_c$, with $L$ the jet half width and $U_c$ its center velocity, so
% that the shown diffusive flux field {\em precedes} the interface.
%
The quantity $\Psi_0$ provides a clear identification of the
turbulent flow region that closely corresponds to the turbulent flow
region marked by the dyed fluid that originally left the jet nozzle.
}
\label{fig.stream0.2}
\label{fig.psi0.2}
\end{figure}
%--------------------------------------------------------------------

\subsection{Barriers of diffusive momentum transport}
\label{sec:psi}
In the Navier-Stokes equations the diffusive momentum transport of an
incompressible fluid is expressed by the term $\vec{h} = \nu \nabla^2
\vec{u}$, with $\nu$ the kinematic viscosity.  As argued by
\citet{Haller2020}, the field $\vec{h}(\vec{x}, t)$ is objective.
While $\vec{h}$ represents the diffusive {\em flow} of momentum, its
{\em flux} involves a surface $A$ with surface normal field
$\vec{n}$.  As time evolves, not only $\vec{h}$ changes, but also the
surface $A$ and the surface normal field $\vec{n}$ are carried along with
the fluid parcels. Using an elementary result of mechanics
\citep{Gurtin2013}, the infinitesimal contribution to the flux at
time $t$ is related to that at time $t_0$ through 
\be
   \vec{h}(\vec{x}, t) \cdot \vec{n} \: \d A = 
   {\rm det} \left[ \nabla \mat{F}_{t_0}^t \right] \;
   \underbrace{ \left[ \nabla \mat{F}_{t_0}^t \right]^{-\dagger}
   \vec{h}(\mat{F}_{t_0}^t(\vec{x}_0), t)}_{\vec{b}_{t_0}^t(\vec{x}_0)}
   \cdot \vec{n}_0 \: \d A_0,
\ee
where $\d A_0$ and $\vec{n}_0$ are the infinitesimal surface area and
its normal, respectively, at an initial time $t_0$.  
In incompressible
3D flow ${\rm det} \left[ \nabla \mat{F}_{t_0}^t \right] = 1$, which
we also adopt for convenience, although we only have 2D information.
The vector $\vec{b}_{t_0}^{t}$ embodies the time dependence of the
flux contribution of fluid parcels that start at $\vec{x}_0$ at time
$t_0$ and flow through the infinitesimal surface that has evolved
from time $t_0$ to time $t$.
The flux contribution, averaged over the Lagrangian path that is
traveled from $t = t_0$ to $t = t_0 + T$, then involves the time
averaged vector $\overline{\vec{b}}_{t_0}^{t_0 + T}$, 
\benn
   \overline{\vec{b}}_{t_0}^{t_0 + T} = \frac{1}{|T|}
   \int_{0}^{T} \vec{b}_{t_0}^{t_0 + t'} \d t'.
\eenn
The vector field $\overline{\vec{b}}_{t_0}^{t_0 + T}$ can be defined
for both forward $(T>0)$ and backward $(T < 0)$ times.  Surfaces that block
diffusive momentum transport come with streamlines of
$\overline{\vec{b}}_{t_0}^{t_0 + T}$ that are tangent to them.
Conversely, the convergence or divergence of streamlines of
$\overline{\vec{b}}_{t_0}^{t_0 + T}$ delineates barriers of diffusive
momentum transport.  These properties can be found from the gradient
field $\nabla \overline{\vec{b}}_{t_0}^{t_0 + T}$ in much the same
fashion as was discussed in Sec.\ \ref{sec:ftle}; technicalities are
detailed below in Sec.\ \ref{sec.res.psi}. The associated field
%%% will be 
is called $\Psi_T(\vec{x}, t)$.  
Summarizing, the vector field $\overline{\vec{b}}_{t_0}^{t_0 + T}$ is
the natural way to express the time-averaged flux of diffusive
momentum.  Its properties can be studied directly, e.g. through its
streamlines, or $\overline{\vec{b}}_{t_0}^{t_0 + T}$ can be used to
measure the flux through turbulent interfaces. 

It is possible to characterize in the same way the {\em
instantaneous} flux, $\vec{b}_{t_0}^{t_0} = \nu \nabla^2
\vec{u}(\vec{x}_0, t_0)$.  
As Fig.\ \ref{fig.stream0.2} shows, the field $\Psi_0$ sharply
defines the boundary between the turbulent and the irrotational
domains, even more acutely than the vorticity field $\omega_z$ (as
shown in Fig.\ \ref{fig.evolution}).  This is remarkable as the
diffusive momentum flow $\vec{h}$ is trivially related to the
vorticity field: in incompressible 2D flow, i.e. $\nabla\cdot \vec{u} =
0$, $\vec{h}$ is explicitly given by $\omega_z$, $\vec{h} = \nu
\nabla^2 \vec{u} \equiv \nu (-\partial \omega_z / \partial y,
\partial \omega_z / \partial x)$. 
%
% \nonote{Lyke: this is the most beautiful field, we have two more
% different kinds, rather than striving for balance, I would go for
% beauty}

While ridge-like local maxima of $\Lambda_T$ are barriers of
large-scale flow, 
%%%
the
%%%
structures $\Psi_T$ emphasize the diffusive flux of
momentum.
The vector field $\overline{\vec{b}}_{t_0}^{t_0 + T}(\vec{x})$ can be
computed from a measured velocity field in much the same fashion as
the finite-time Lyapunov field.  While $\Lambda_T$ involves the
gradient velocity field $\mat{A}$, a measurement of $\nabla^2
\vec{u}$ takes one more derivative.  It is regularized by averaging $\overline{\vec{b}_{t_0}^{t}}$
over a time $T$, but the gradient matrix $\nabla \vec{F}_{t_0}^t$,
which strongly fluctuates along Lagrangian paths, now adds to the
noise.
However, turning the vector field $\overline{\vec{b}}_{t_0}^{t_0 +
T}$ into $\Psi_T$ significantly enhances its signal to noise ratio.

\begin{figure}
\centering
\includegraphics[scale = 0.5]{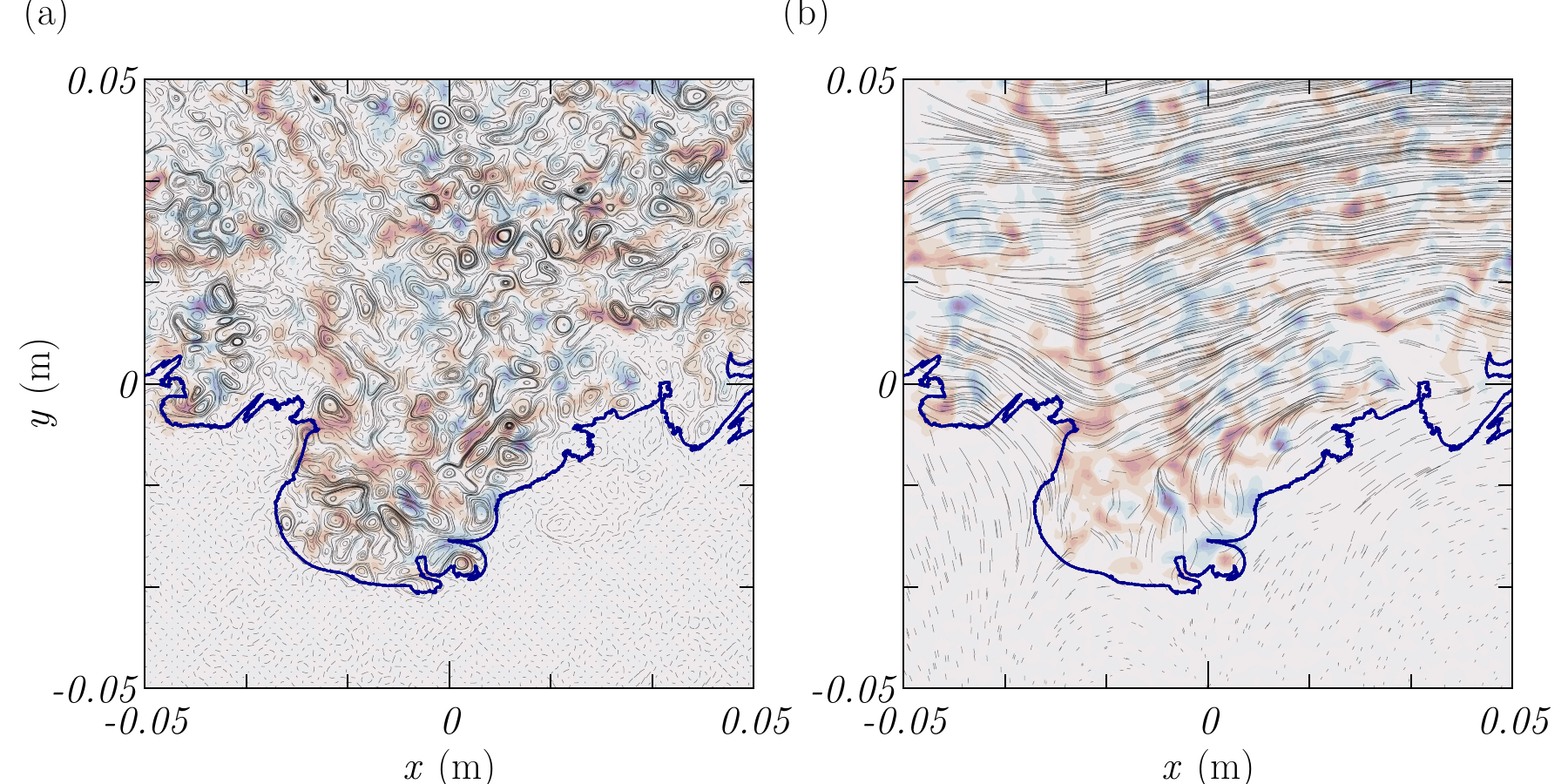}
\caption{ Instantaneous streamlines 
(a) the diffusive momentum flux and
(b) advective transport of tracer particles. The blue line represents the \tnti.}
\label{fig.streamlines}
\end{figure}

Fig.\ \ref{fig.streamlines} illustrates instantaneous transport streamlines resulting from the advective transport of passive tracers ($\vec{u}$) and the diffusive transport of linear momentum ($\nu \nabla^2\vec{u}$). As soon as integration time goes above zero time, these streamlines change in time and become pathlines. The finite-time Lyapunov exponent acts as an operator on these pathlines to identify ridges that define Lagrangian coherent structures. Advective streamlines primarily stretch along the dominant advection direction and also transport across the interface (see Fig.\ \ref{fig.streamlines}). Diffusive streamlines, on the other hand, form small-scale structures irrespective of the main flow direction.
%%

%--------------------------------------------------------------------
%--------------------------------------------------------------------

%.............................................................. Setup
%....................................................................
\section{Experimental setup}
\label{sec:setup}
\begin{figure}[t]
\centering
 \includegraphics[width=1\linewidth]{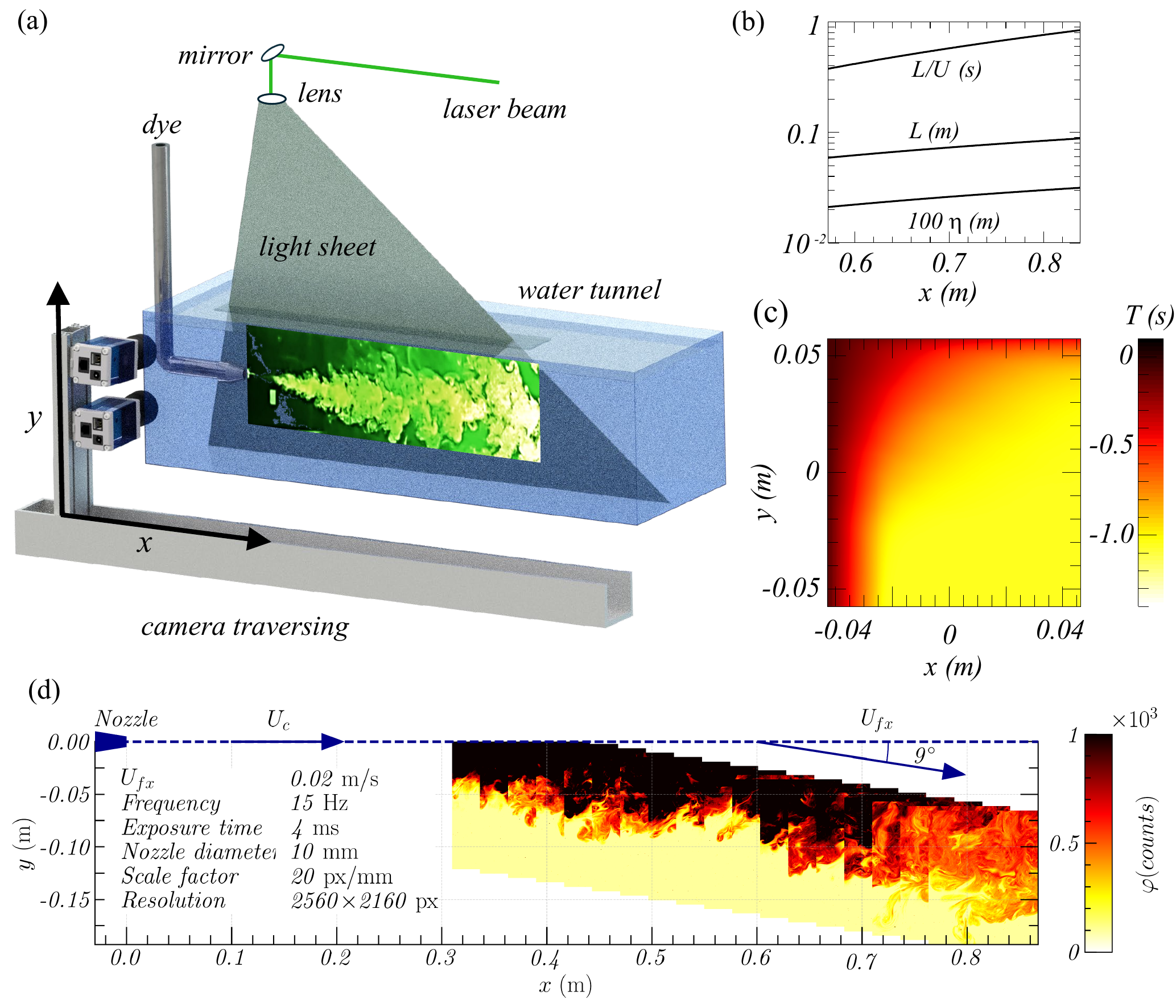}

\caption
{
Experiment characteristics.  (a) The PIV and LIF cameras move with
the \tnti\ interface.  The horizontal component of the frame traverse
velocity $\vec{U}_{fx}$ is $2 \: {\rm cm / s}$, and points $9^\circ$
downward.  
(b) Variation over the region of interest of relevant length and time scales, where  
$L$ is the local jet half-width,  $U_c$ the local mean velocity at the jet centerline, and $\eta$ the local Kolmogorov scale; 
%%% The local integral time scale is 
$L/U_c$ is the local integral time scale. %with $U_c$ the jet centre velocity.  
%%% The values are taken for the jet centre line.
%
(c) Average residence time of fluid parcels within the field-of-view
% in a typical co-moving frame
%%% when tracing them 
trajectories backwards in time (i.e., $T < 0)$. 
In the upper half of the frame, the observation time is limited by the
higher velocities towards the 
jet centerline, while on the left side of the frame, quiescent fluid outside the jet 
exits the field-of-view during the motion of the cameras.
(d) The concentration fields of a single run as a function of global
coordinates.
}
\label{fig.setup}
\end{figure}

%
%\note{Ali text (nice!)}
%
% \marginpar{We need a diagram such as Fig. 2.2 in Coen's M.Sc. thesis}
Water seeded with 
%%%
a fluorescent dye (rhodamine-6G)
% \marginpar{rhodamine-6G; see Coen's report}
%%% dye 
flows through a 10~mm diameter jet nozzle with
nominal velocity of 1.25~m/s. The flow 
%%%
through the nozzle is controlled
%%%.
and emanates in the water-filled test section of a water channel with a 0.60$\times$0.60~m$^2$ cross section and a length of 5.0~m.
%%%
The jet
%%% resulting 
% corresponding 
%%%
Reynolds number is ${\rm Re}$ = (1.25$\pm$0.03)$\times$10$^4$, which is above the mixing transition (${\rm Re} \approx 10^4$)
\citep{Dimotakis2000}. 
% \marginpar{Here you can mention the Kolmogorov scale and the Taylor scale, and provide a ref on how the dissipastion rate was estimated}
%%%
The dissipation rate $\varepsilon$ is estimated from $\varepsilon$ = 0.015$U_c^3/L$ \cite{Panchapakesan93}, where $U_c$ is the local mean centerline velocity of the jet, and $L$ the jet half-width. This gives a Kolmogorov length scale of $\eta$ = 0.20~mm at the start of the measurement (at $x = 0.53$~m from the nozzle exit), and a Taylor length scale of $\lambda_T$ = 4.9~mm.
% Taylor length scale is estmated from 
% \varepsilon = 15\nu u'^2/\lamba^2 => eps = 0.676e-3 m^2/s^3
% nu = 1e-6 m^2/s
% u' = 0.25 U_c, U_c = 0.13 m/s
% \lambda = \sqrt{15\nu/varepsilon} u' = 4.9 mm [check]
%%%
The detection of concentration and velocity is
designed to move with the \tnti.  The $x,y$-traverse system is
driven by two stepper motors (MDrive23Hybrid, Schneider Electric,
USA), with a 2~m span along the $x$-axis parallel to the jet axis,
and a 1~m span in the $y$-direction.  
The jet characteristics vary with the distance $x$ to the nozzle, and
are detailed in Fig.\ \ref{fig.setup}(b). 
%
%\note{PLEASE provide details: exposure time, frame straddling}
The flow in the test section is illuminated with a thin laser light sheet with a thickness of 1.5~mm, generated from a dual pulsed Nd:YAG laser (Spectra-Physics PIV-400).
% \marginpar{What is $\Delta t$ in the measurements?}
Detection of the 
%%%
dye
%%%
concentration (using LIF) and 
%%%
flow
%%%
velocity (using PIV)
fields involve two high-resolution 
%%% LaVision Imager 
sCMOS 
% CLHS 
cameras 
%%%
(LaVision Imager CLHS)
%%% CLHS stands for CamLink HS}
operating at a framing rate of 15~Hz, with the double frames separated by 4~ms exposure time. The LIF image corresponds to the first PIV frame. 
The LIF camera is positioned 0.1~m above the PIV camera and tilted downwards by 3~degrees to match the field of view of the PIV camera (see Fig.\ \ref{fig.setup}(a)). 
We mount a long-pass filter (SCHOTT OG570) on the LIF camera and a 525~nm bandpass filter (TECHSPEC) on the PIV camera. Both cameras use 105~mm lenses. 
We use a calibration grid and an image mapping function to overlay
the two measurement fields with an error of less than 0.2 pixels.

% \marginpar{add make and model of cMOS cameras}

% \marginpar{There are 2 LIF recordings per PIV image pair; which one do you use?}
%
%\nonote{Exposure time ?}
% Please specify the variation of the nozzle exit velocity over
% the duration of the experiment.  How well could the nozzle be
% identified in our camera fudge ?}
%
The cameras move at a constant velocity of 0.02~m/s and a 9
degree angle with respect to the jet axis as shown in Fig.\ \ref{fig.setup}(d).
However, the interface velocity decreases as a function of the downstream location. This means that at the start of the traverse, the interface velocity is higher, but it becomes lower than the traverse velocity toward the end. As a compromise, we considered only the part of the traverse where both velocities match (see Fig.\ \ref{fig.setup}.(b)).
%
% \marginpar{Mention that the traversing system can only move at a constant velocity and that this is a compromise; what is the jet velocity at $\eta$ = 2 on average over the measurement range?}
%
%%% They 
The cameras have a common 
%%% have a 
field of view of 100$\times$120~mm$^2$ with a scale factor of 0.05~mm/px.  
% \marginpar{This is not the image magnification; you need the image magnification to get the depth-of-field.}
The 
%%% first 
LIF camera 
%%%
only records the
light emitted from the rhodamine dye using an optical long-pass
filter, 
% \marginpar{filter specs?}
while the PIV camera records the light scattered off
spherical hollow glass particles, allowing for simultaneous LIF and
PIV measurements 
%%% at a camera F stop 
with an aperture number $f^\#$
%%%
of 5.6.
% \marginpar{Specify the lenses used; does the PIV camera also record the LIF signal? Or did you use a filter?}
%
A scalar calibration was done to ensure a linear relation between the
dye concentration $\varphi(\vec{x}, t)$ and the observed fluorescence
intensity.  
%%% In the sequel, therefore, we express $\varphi$ in intensity levels.
In the remainder of this paper $\varphi$ is expressed in intensity counts.
% \marginpar{Then why do you calibrate?}

% The PIV and LIF spatial resolutions are $10\:\eta$ and $0.5\:\eta$,
% respectively, where $\eta$ is the Kolmogorov scale.
%

%------------------------------------------------------------------
% \marginpar{details such as framing rate, field of view, velocity and direction of traversing should all go to the text.}
% \marginpar{fig. 3(d) looks square, but should be rectangular with equally scaled axes}
%--------------------------------------------------------------------
% \nonote{Lyke: Fig.2d is a single snapshot.  I will change $U_x, U_y$.
% Indeed, flow coming from right is unusual, but let us keep it (for a
% while) to prevent mistakes.  Flipping will affect a lot of pictures.}

% \note{Where is $\eta$ taken, at which $x$ ?}
\subsection{Analysis of LIF and PIV images}
\label{sec:LIF_PIV_analysis}
The 2105$\times$2563-pixel LIF images are 
%%% averaged 
filtered
%%%
using a
Gaussian width of 4 pixels (standard deviation $(4/2)^{1/2}$ pixels)
% Gaussian kernel with a Gaussian width of 4 pixels,
% \marginpar{why? Is 4~px the e$^{-2}$ width or st.dev.?} 
% \marginpar{check the phrase. There is a broken sentence here}
and 
%%% were
subsequently downsampled to 508$\times$638-pixel images. Filtering reduces photon noise but at the expense of spatial resolution.  
% \marginpar{why?}
The
resulting 
%%% 
equivalent pixel size 
%%%
in the object plane
%%% 
is 2$\times$10$^{-4}$~m, which is
approximately the 
%%%
estimated average 
%%%
value of the Kolmogorov length $\eta$
%= 2\times10^{-4}\; {\rm m}$
at the beginning of the region of interest.
% \marginpar{$x$=?; how did you compute the Kolmogorov scale?}
%%% ε = 0.015U_c^3 /b_u (Panchapakesan & Lumley 1993)
% U_n = 1.25 m/s; D = 0.010 m; B = 6.0; z = 0.58 m.
% U_c = B U_n D / z = 0.13 m/s
% b_u = 0.082 z = 0.048
% eps = 0.676e-3 m^2/s^3
% eta = 0.20 mm
It should be compared to
the 
%%% $1.5\times 10^{-3} \: {\rm m}$ 
1.5~mm width of the light sheet and the vector spacing $\approx 10^{-3} {\rm m}$ of the measured velocity field.
% \marginpar{The thin features may not always be averaged over the width of the light sheet. This should be rephrased. Also compare with the Taylor microscale}
%
% Done in kde.for
%
Occasionally, dust particles 
%%% , clad by dye,
% \marginpar{we are not sure that the dye attached to the particles; it could also be fluorescent dye reflected off the particles}
may light up 
or fluorescent dye reflected off the particles
%%% brightly 
in the LIF 
images, which causes cluster analysis to fail. These spots, 
leading to isolated peaks at the highest intensity in histograms,
were removed from the images using a median filter and replaced by an average over
background pixels. %%% put this in the main text:
The results in this paper are from 200 images in each run,
which span the 
%%%
0.57$\times$0.84~m$^2$
region of interest,
%%% $x \in [0.57 \: {\rm m}, 0.84 \:{\rm m}]$, 
with $x$ the distance from the 
%%% $1 \; {\rm cm}$ diameter
nozzle, and taken with a frame rate of 15~Hz.  Averages are
over 15 repeated runs.\\
%%% END put this in the main text
% \marginpar{How did you remove them? minmax filter?}

Two-dimensional sections of the velocity field are measured using a
multigrid PIV algorithm with rectangular interrogation windows
\citep{Adrian2011}, tailored to the large variation of the fluid
velocity over the region of interest. 
The initial size of the interrogation regions is $256 \times 64$-pixel, and the final square size $32 \times 32$-pixel with $50 \%$ overlap.
The velocity field from PIV is
finally evaluated on a grid of 113$\times$145 interrogations, with a
%%% vector 
spacing of 7.7$\times$10$^{-4}$~m. %%% move this to the main text
The frame
velocity ($U_{f x}$ = 2$\times$10$^{-2}$~m/s, $U_{f y} =
-$3.2$\times$10$^{-3}$~m/s) is added the the flow velocity; see
Sec.~\ref{sec:move}).  
%
% \marginpar{What is the initial size of the interrogation regions in px, and the final size in px, and how much is the overlap? Is the final interrogation window size square? Specify also the spatial resolution as twice (?) the vector spacing, assuming that the final interrogation window is square; compare to the Kolmogorov and Taylor scales and the light-sheet thickness}

\subsection{Moving with the flow}
\label{sec:move}
\begin{figure}[t]
\centering
 \includegraphics[width=0.9\linewidth]{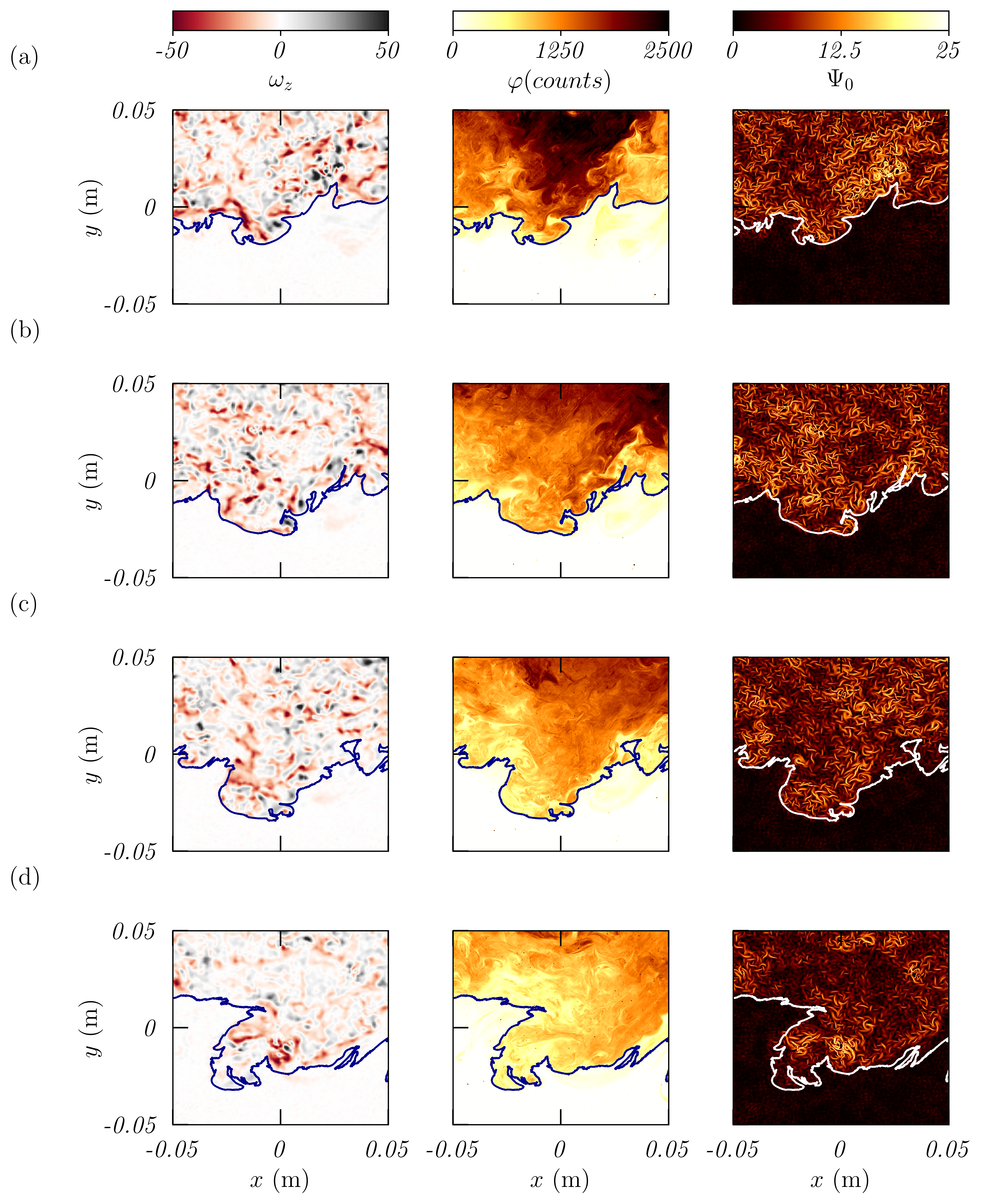}
\caption
{Quasi-Lagrangian evolution of flow structures at the \tnti\. (a) to (d) sequential evolution of an engulfment event quantified with vorticity $\omega_z$, scalar $\varphi$, and diffusive barrier $\Psi_0$. }
\label{fig.evolution}
\end{figure}

Moving at the interface's average velocity allows the evolution of flow structures within the field of view to be frozen, in contrast to stationary measurements where structures enter and exit the field of view (see figure~\ref{fig.evolution}).
There are several ways to interpret experiments in which the
detection involves a moving frame of reference.  They affect the appearance of
Lagrangian tracks $\vec{x}(t)$, which in the laboratory frame follow
from 
\be
   \frac{\d \vec{x}}{\d t} = \vec{u}(\vec{x}, t).
\label{eq.move.1}   
\ee   
Let $\vec{x}$ and $\vec{u}(\vec{x}, t)$ be the position and velocity,
respectively, in the laboratory frame, and $\vec{x}'$ and
$\vec{u}'(\vec{x}', t)$ those in the moving frame. 
%%% The
In the present case 
frame moves
with
a constant 
velocity $\vec{U}_f$, so that $\vec{x}' = \vec{x} - \vec{U}_f
t$.

One way, as is done here, is to add the frame velocity to
the velocity in the laboratory frame, 
\be
   \vec{u}'(\vec{x}', t) = \vec{u}(\vec{x}', t) + \vec{U}_f,
\ee
so that $\vec{x}' = \vec{x}$. Lagrangian 
%%% tracks, 
trajectories, 
measured in the
moving frame, then follow from
\be
   \frac{\d \vec{x}'}{\d t} = \vec{u}'(\vec{x}', t) - \vec{U}_f,
\ee
with $\vec{u}' = \vec{u} + \vec{U}_f$; this is exactly the equation in
the laboratory frame.
%
%\note{which must be checked...}
%
A stationary fluid parcel in the laboratory frame now also has zero
velocity in the moving frame.  
%
%%% The other 
An alternative
interpretation is to use the information in the moving
frame `as is', but then the Lagrangian 
%%% tracks 
trajectories 
no longer represent
those in the laboratory frame.

We trace fluid parcels backward in time ($T < 0$).  The observation
time in the moving frame is shown in Fig.\ \ref{fig.setup}(d).  A
small velocity of fluid parcels at the very left edge of the moving
frame 
%%% lets them move out quickly, 
is the cause 
%%% so 
that their observation time
$|T|$ is small.  It is also small 
%%% in 
near 
the core of the jet, and it is
large at the interface location.
The third component of the velocity, which corresponds to out-of-plane motion, is estimated to move particles away from the field of view after 1.5 local integral time at the beginning of the traverse and 0.8 local integral time at the end, at the interface location.

%------------------------------------------------------------------
% \note{Add colorbar \& caption}

%--------------------------------------------------------------------

% %\note{This paragraph will be removed}
% %
% The other way, followed in \citep{Jesse2021}, is to just work with
% the velocities from PIV on the moving frames, without correction for
% the frame velocities, so that $\vec{u}'(\vec{x}', t) =
% \vec{u}(\vec{x}, t) - \vec{U}$.
% %
% \note{This alinea goes out.  We never worried about this.  In
% \cite{Jesse2021} we called this the ``moving-frame ftle field'',
% which is ok and unambiguous.}
%
% The track equation in the moving frame is now
% \be
%    \frac{\d \vec{x}'}{\d t} = \frac{\d \vec{x}}{\d t} - \vec{U} =
%    \vec{u}(\vec{x}, t) - \vec{U} = \vec{u}'(\vec{x}', t),
% \ee
% which is the same as Eq.\ (\ref{eq.move.1}), but tracks differ from
% those in the laboratory frame as $\vec{x} = \vec{x}' + \vec{U} t$.

%...................................................... Images mania
%
%%% \section{Finding the turbulent interfaces}
\section{Identifying the turbulent interfaces}
\label{sec:find.tnti}
The demarcation between zero and finite values of the enstrophy (the
\tnti) requires a threshold value $\omega_{\rm thr}^2$ of the
enstrophy.  In the simulations of \citet{Vassilicos2023}, who took
care of numerical oscillations in the region of quiescent fluid, the
interface location was found insensitive to $\omega_{\rm thr}^2$ over
a large dynamic range ($ 10^{-8} \lesssim \omega^2_{\rm thr}
\lesssim 10^{-2}$).  In the context of experiments, inevitably
influenced by noise, the determination of a threshold is more
ambiguous, and the threshold levels are not robust.

\nonote{It is more than proxy, it is a different field with different
dynamics.  Still all papers do the edge of $\varphi$ is {\em the}
interface.}
The measured scalar concentration field $\varphi(\vec{x}, t)$, i.e.
the observed fluorescence, is taken as a proxy of the vorticity, 
which in two dimensions satisfies the same equation as
$\varphi(\vec{x}, t)$, but with a 
%%% different diffusivity.
negligible diffusivity, so that the dye can effectively be considered as a passive tracer that follows the motion of the fluid elements that passed through the nozzle.\footnote{For the dye used in this experiment (rhodamine 6G, $\mathbb{D}$ = 2.8$\times$10$^{-10}$~m$^2$/s), the
Schmidt number ${\rm Sc} = \nu / \mathbb{D}$, with $\nu$ the
kinematic viscosity and $\mathbb{D}$ the molecular diffusivity, is
Sc = 3.6$\times$10$^3$. 
%{\cal O}(8\times 10^3)$ \cite{Crimaldi2008}.  
Therefore,
the Batchelor scale $\eta_B=\eta/\sqrt{\rm Sc}\cong$ 3-5$\times$10$^{-6}$~m
remains unresolved. 
% , and what matters is the ratio between the {\em turbulent} diffusivities of mass and momentum which is $\approx 2$ \citep{Townsend1976}.
} 
%
% \nonote{check the Schmidt}
%%% The molecular diffusivity of rhodamine-B in water is 3.6E-10 [m^2/s]. This gives a Schmidt number of 2.8E3. For rhodamine-6G the diffusivity in water is 2.8E-10 [m^2/s], which gives a Schmidt number of 3.6E3. In the report of Coen it is mentioned that rhodamine-6G was used.
%%% DOI 10.1007/s10895-008-0357-7
%%% NB I checked with Lyke and Edwin, and they confirmed Coen used R6G (most likely guess)

We illustrate the 
%%% quest 
identification
of the \tnti\ based on scalar concentration.
Ideally there is dye in the seeded turbulent jet, and no dye outside
(the `blue sky'). However, after repeated runs the region of
unmixed fluid may become contaminated by a low background concentration
$\varphi_{\rm bg}$.  Then, the \tnti\ is the boundary between the
region with $\varphi_{\rm bg}$ and the domain with larger
concentration. 
%%% Ref. to Buxton ??

\begin{figure}[t]
\centering
 \includegraphics[width=1\linewidth]{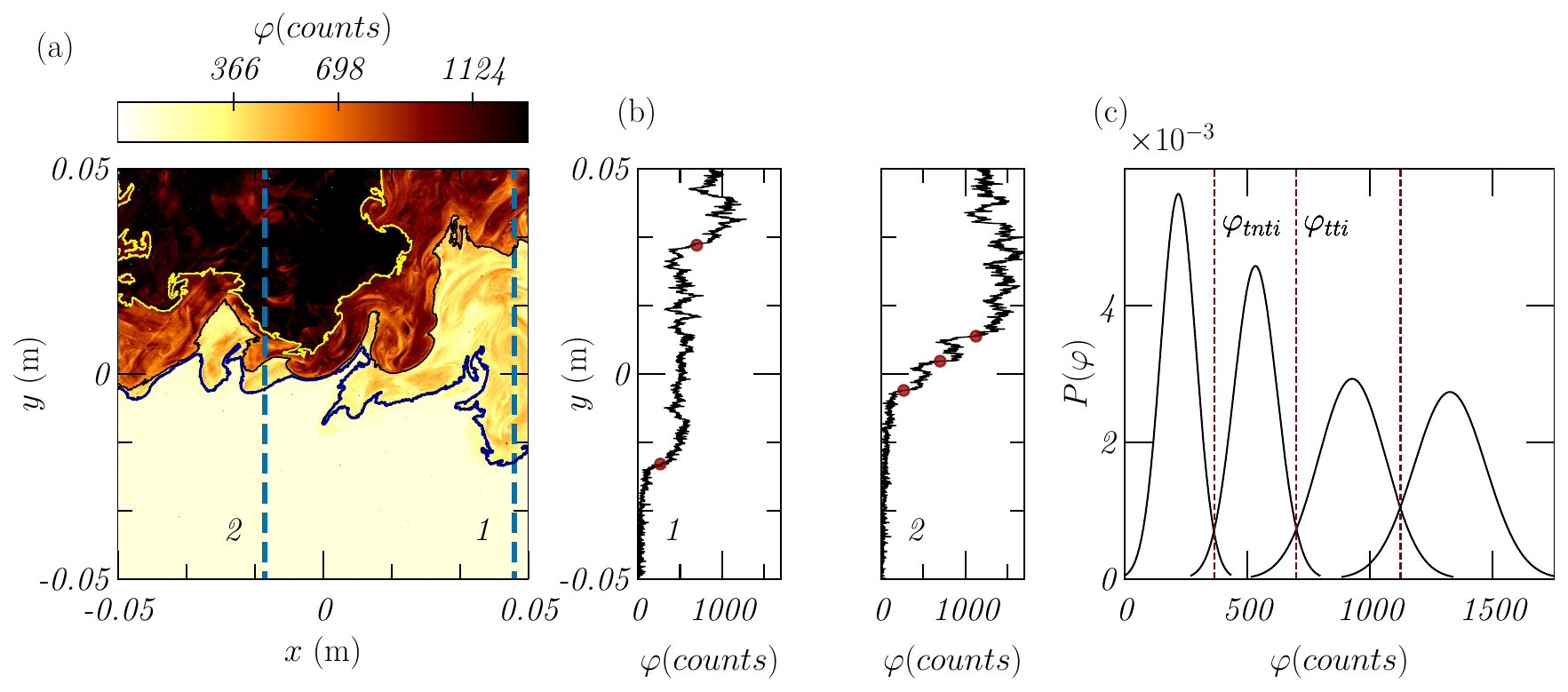}
\caption{
%
%%% Finding 
Detecting the \tnti\ using cluster analysis.  (a) Fluorescence image intensity
$\varphi(\vec{x}, t)$ (expressed in pixel counts). The lower side of the black line is the \tnti, the upper
side is a \tti.
(b) Concentration profiles along the blue vertical lines (1,2) in % frame
(a).  The red dots indicate the contour concentration values
$\varphi_{\rm tnti}$ and $\varphi_{\rm tti}$ that define the \tnti\ and \tti,
respectively.
Concentration values are subtracted from the background and shifted to have a nonturbulent part starting from zero. 
(c) Cluster distributions of the pixel intensity levels corresponding to the 4 uniform concentration zones.
The red dashed lines indicate the intersection $\varphi_{\rm tnti}$ between
the first and second clusters, and $\varphi_{\rm tti}$ between the
second and third clusters.  Contours at the corresponding
concentration values are drawn in (a).
A second \tti\ is also drawn in (a).  Its concentration level
follows from the intersection of the third and fourth clusters.
}
\label{fig.cum}
\end{figure}

It appears that in 
%%%In 
a turbulent jet flow seeded with dye there are more distinct
concentration levels than just these two.  
%%% Ref to Buxton and others?
The scalar concentration
field appears to be organised in uniform concentration zones, i.e. regions
where the concentration variation is small~\citep{Dahm1987}. 
%%%
Various approaches exist to identify the boundaries between these regions; here these
%%% These
uniform concentration zones 
%%% may be found 
are identified 
by cluster analysis.  This
well-established statistical technique \citep{Bezdek1980} arranges
the pixels containing concentration values into clusters,
%%%
as illustrated in Fig.\ \ref{fig.cum}.
%%% It was first used 
\citet{Fan2019} originally used this approach 
to find turbulent interfaces. 
The optimization procedure
uses no spatial information. Although the choice of the number of
clusters $n_c$ can be done automatically \citep{Silverman1986}, we
%%% have 
take 
% \marginpar{why 4?}
$n_c = 4$; 
this is the minimum number of clusters to include the identification of the ambient fluid and up to 3 internal uniform concentration zones.
%%% The method is illustrated in figure~\ref{fig.cum}.  

Each cluster has its own 
%%% PDF, 
concentration distribution; 
these are shown in Fig.\ \ref{fig.cum}(c).  The concentration level of an interface is taken
as the intersection between the corresponding 
%%% PDF's
distributions for each cluster.  For the \tnti\
and the \tti\ these concentration values are $\varphi_{\rm tnti}$ and
$\varphi_{\rm tti}$, respectively.  Finally, the interfaces are drawn
as contours at these concentration values.  The essence of the
clustering algorithm is an optimal association of concentration
values with a 
%%% few 
small number of
clusters.  As Fig.\ \ref{fig.cum}(c) illustrates,
these associations may overlap.  
An interface marks a jump of the concentration value; those jumps are
illustrated in Fig.\ \ref{fig.cum}(b).  
% \marginpar{what do you mean with this sentence?}
Clearly, identifying those
jumps is aided essentially by the clustering algorithm. 

Closed contour loops, which correspond to patches of dye that appear
unconnected to the turbulent domain, and loops encircling patches of
irrotational fluid inside turbulence were removed.  In the used
contouring algorithm \citep{Bourke1987}, contours come in pieces, of
which we kept the 6 longest ones.  
%
%%% Our detection moves with the (average) velocity of the edge of the
%%% jet, which is much smaller than the 
%%% center 
%%%velocity at the jet centerline. 
%%% Therefore, dye may be encountered that was deposited before the instant of observation: detrained scalar. It was verified that these patches do not contain vorticity.  
% \marginpar{altenative}
Occasionally, patches of dye are found on the irrotational side to the \tnti; this is jet fluid that was detrained before the instant of observation, since the camera, that on average follows the edge of the jet, moves at a velocity that is much smaller than the core region of the jet.
It is verified that these patches do not contain vorticity.
% \nonote{Ali will verify this...}

%--------------------------------------------------------------------
%--------------------------------------------------------------------

Our method should be compared to the popular procedure in which the
contour level $\varphi_{\rm tnti}$ is determined from an inflection
point in the 
%%% cumulant of the concentration PDF 
%%% cumulant means something very different in statistics!
cumulative distribution of the pixel intensities
\citep{Prasad1989}. We
found that this 
%%% concentration value could not always be found unambiguously. 
approach does not always yield an unambiguous threshold value.
In contrast, once the number of clusters $n_c$ is set,
this ambiguity is no longer present in our method. 
Clusters increasingly overlap with increasing $n_c$.  The first two
%%% PDF's 
intensity distributions 
in Fig.\ \ref{fig.cum}(c) are well separated
%%%,
%(at threshold values of 900 and 1400 counts, respectively),
%%%
but the
definition of the second \tti\ in Fig.\ \ref{fig.cum}(a) 
%%%
% \marginpar(check value)
%(at a threshold value of 1880 counts) 
is less
acute (Threshold values of clusters are shown as tickers on the colorbar). 
% 
% Further, it is not obvious how to extend the cumulant method to the
% \tti.

%....................................................................
% \subsection{Internal layers}
%
Cluster analysis of the concentration field provides a natural way to
find turbulence interfaces as the edges of clusters. Internal
turbulence interfaces in a turbulent boundary layer were studied by
\citet{Eisma2015pof}.  Their detection required the distinction of
turbulence levels, which was done on the basis of the shear vorticity
\citep{Kolar2007}.  
Using the scalar field, turbulent--turbulent interfaces 
%%% have been
were
studied by \citet{Buxton2023}, with a dyed turbulent wake evolving in
background turbulence.  In their case, the \tti\ is the interface
between turbulence with dye, and turbulence without dye.  In
contrast, our \tti\ is the interface between two non-zero
concentration levels.  Below we present conditional averages of
$\omega_z, \Lambda$, and $\Psi$ both on the \tnti\ and on this \tti.

% \nonote{We could refer to Masoud Asadi, but his method paper is not
% (yet) published:\\ Using the PDF's of velocity and vorticity, together
% with judiciously chosen threshold values, they identify a \tti\ of
% tbl in freestream turbulence.}

%....................................................................
\section{Conditional averages}
\label{sec:cond}
The turbulent interfaces are found from the measured dye
concentration. To establish the relation with the scalar fields
$\omega_z(\vec{x}, t), \Lambda_{-T}(\vec{x}, t), \Psi_{-T}(\vec{x},
t)$, and $\Psi_0(\vec{x}, t)$, we use the conditional average as
presented by \citet{Bisset2002}.
The question is whether structures of a scalar quantity are aligned
with an interface.  If so, the conditional average of the scalar
should vary sharply at this interface, and should be structureless
anywhere else.  

The conditional average $\wt{\omega_z}(s)$ of the vorticity component
$\omega_z$ with respect to the \tnti\ location $\vec{x}_{\rm tnti}$
is defined as
\benn
   \wt{\omega_z}(s) = 
   \left\langle \omega_z(\vec{x}_{\rm tnti} + s \vec{e}_{\rm tnti})
\right\rangle_{\svec{x}_{\rm tnti}}, 
\eenn
with the unit vector $\vec{e}_{\rm tnti}$ directed perpendicularly to
the \tnti\ $\vec{x}_{\rm tnti}$, and where $s < 0$ is the
irrotational domain while $s \: \vec{e}_{\rm tnti}$ with $s > 0$
points into the turbulent domain. In this paper, the tilde symbol (e.g., $\wt{\omega}$) denotes a scaled, non-dimensionalized quantity. Averages 
%%% $< \;\; >$ 
$\langle\cdots\rangle$
are done over
the interface, and over all frames and all experiment runs.  The
conditional average of other scalar quantities is defined
analogously.  The experimental 
%%% information 
data 
is not free of noise so
that $\wt{\omega_z}(s < 0)$ would still be finite, and even more so
for $\wt{\Psi_{-T}}$ and $\wt{\Psi_0}$. 
\nonote{Only for vertical intersections $\wt{\omega_z}(s)$ the
average is over lines {\em parallel} to the interface, as in
\cite{Westerweel2005}.}

\begin{figure}[t]
\centering
\includegraphics[scale = 0.9]{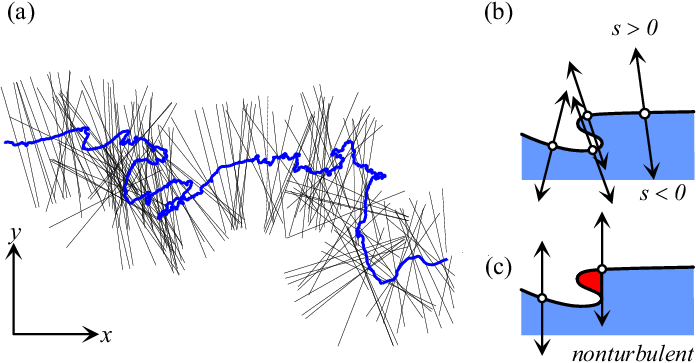}
\caption{Conditional averages.  
(a) The blue curve represents a \tnti, black lines: edge normals $s
\vec{e}_{\rm tnti}$. The interface folds back on itself:
(b) intersections for edge normal $s \vec{e}_{\rm tnti}$;
(c) vertical intersections $s \vec{e}_{y}$ for \tnti\ outer and inner envelopes, which either include or exclude the red
patch of nonturbulent fluid.
}
\label{fig.normal}
\end{figure}

We define an interface normal intersection $\vec{e}_{\rm tnti}$ by
first fitting lines to $\vec{x}_{\rm tnti}$ with length $\Delta =
16\: \eta$.  Edge normals $\vec{e}_{\rm tnti}$ are the bisectors of
these lines.  A typical result for the normals $\vec{e}_{\rm tnti}$
on the \tnti\ is shown in Fig.\ \ref{fig.normal}(b).  In choosing
edge normals, the interface is covered with boxes with size $\ell_{\rm
box} = 16 \: \eta$, with one intersecting point in each nonempty box.
Consequently, the density of edge normals is high where the interface
is very contorted, which biases conditional averages. 
This procedure respects the fractal character of the interface shape.
%
%\nonote{The length of these line fits should also scale with $x$ !}
%
It is well known that the interface has fractal properties  
\citep{Prasad1989,Meneveau1990,Mistry2016,Mistry2018,Vassilicos2023,Buxton2023},
with the number of nonempty boxes diverging faster than $\ell_{\rm
box}^{-1}$ with decreasing box size $\ell_{\rm box}$.

An alternative choice of intersections is to take the unit vector
$\vec{e}_{\rm tnti}$ in the {\em vertical} direction $\vec{e}_y$ 
%%% as was done by 
\cite{Westerweel2005}.  This approach needs a procedure
to deal with sections where the interface folds back on itself, as illustrated in Fig.\ \ref{fig.normal}(b,c), and where the \tnti\ is
chosen as 
%%% 
either the outer or inner 
envelope of the nonturbulent domain.  The choice made
-- to 
%%% exclude 
include %%% ??
engulfed irrotational fluid -- introduces a bias in the
conditional average.  In a few cases in Sec.\ \ref{sec:results.omega}
we demonstrate the effect of these choices and find that this bias is
small most of the time.  
%
%--------------------------------------------------------------------
%--------------------------------------------------------------------

Taking perpendicular cross sections along $\vec{e}_{\rm tnti}$
presents a challenge for very contorted interfaces.  A few examples
are sketched in Fig.\ \ref{fig.normal}(c), where the line $s \:
\vec{e}_{\rm tnti}$ may intersect the turbulent domain several times.
If $s$ is the coordinate along the intersection, it is assured that
the turbulent domain corresponds to $s > 0$ and $s < 0$ is the region
of unmixed fluid.  This is done by computing the integrated scalar
concentration 
$\varphi^+ = \int_0^{+\ell/2} \varphi(\vec{x}_{\rm tnti} + s \:
\vec{e}_{\rm tnti}) \: \d s$, and similarly for $\varphi^-$, and
choosing the sign of $s$ such that $\varphi^+ > \varphi^-$.
Intersecting lines with $| \varphi^+ - \varphi^- | / (\varphi^+ +
\varphi^-) < 0.1$ are deemed ambiguous and excluded from the
conditional average. Before averaging individual sections, the
coordinate $s$ is scaled with the jet half width, whose variation
during a run is shown in Fig.\ \ref{fig.setup}(c).
In the case of multiple intersections further refinements are
possible, such as only including intervals of $s$, $s > 0$, in the
conditional average that actually correspond to the turbulent domain,
and {\em vice versa} for the irrotational region.  These refinements do not
significantly change our results.
%
%\nonote{There is a figure proving this.}
%
Conditional averages with respect to the \tti\ are done analogously:
the sign of $s$ is again chosen such that $\varphi^+ > \varphi^-$.
% A sermon...
% The result of an average conditioned on the \tnti\ depends on the way
% it is taken - be it with perpendicular or vertical intersections. For
% perpendicular averages it also depends on the contortions of the
% interface, which determines the local density of intersections.

%............................................................ Results

%....................................................................
\section{Results}
\label{sec:results}
\subsection{Conditional averages of $\omega_z$}
\label{sec:results.omega}
\begin{figure}[t]
\centering
\includegraphics[scale = 0.9]{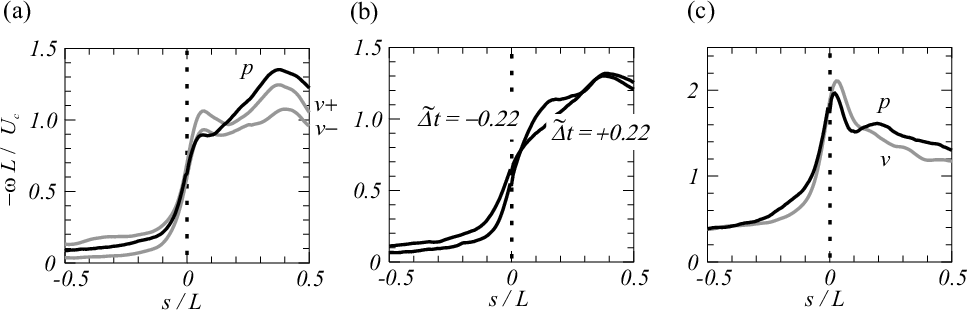}
\caption{Conditional averages of the out-of-plane component $\omega_z$ of the vorticity.  
(a) Averages over normal ($p$) and vertical ($v_\pm$)
intersections of the \tnti.  The $v_\pm$ averages distinguish two
envelopes: $v_-$ is the envelope of the nonturbulent domain, as is
illustrated in Fig.\ \ref{fig.normal}, while $v_+$ is the envelope of
the turbulent domain.
(b) Influence of the time delay $\Delta t$ between the scalar fields
$\varphi(\vec{x}, t)$ and $\omega(\vec{x}, t + \Delta t)$, and thus
between perpendicular intersections of the \tnti\ at $t$, and
$\omega_z$ at $t + \Delta t$.  The time shift $\wt{\Delta t}$ is
expressed in units of $L / U_c$ (at $x$ = 0.71~m).  A delay
$\wt{\Delta t}$ then corresponds to two frames.
(c) Conditional average with respect to the \tti\ (notice the change
of the vertical scale).
}
\label{fig.res.omega}
\end{figure}

Fig.\ \ref{fig.res.omega} shows conditional averages of $\omega_z$, where
Fig.\ \ref{fig.res.omega}(a) illustrates the two choices of
conditional averages on vertical intersections of the \tnti.  Both
choices either take the envelope of the turbulent domain or take the
envelope of the nonturbulent region result in different conditional
averages.  They both differ from the 
%%% average that is done 
conditional averages 
along
(perpendicular) edge normals.  In the 
%%% sequel, 
remainder of this paper, we take
%%% chosen 
averages along the proper edge normals of the turbulent
interfaces. The result in Fig.\ \ref{fig.res.omega} can be compared
to the result of \citet{Mistry2016}, which is for $\wt{|\omega_z|}$ at
a Reynolds number e = 2.5$\times$10$^4$.
\nonote{Mistry has poor(er) resolution, AND absolute values. So only a
crude comparison is possible.  Lyke: maybe there is no comparing
possible.}

%--------------------------------------------------------------------
%--------------------------------------------------------------------

% \marginpar{$\omega_z$ is determined at $t+\frac12\Delta t$, while the concentration is determined at two consecutive times. We should have used this to determine $E_b$, but there is no time for that. Also, where does the interrogation put the velocity measurement using window shifting?}
The conditional average of $\omega_z$ in Fig.\ \ref{fig.res.omega}(b)
depends on the time delay $\Delta t$ between the scalar fields
$\varphi(\vec{x}, t)$ and $\omega_z(\vec{x}, t + \Delta t)$, and thus
on the time delay between the \tnti\ at $t$ and $\omega_z$ at $t +
\Delta t$.  The jump of $\wt{\omega_z}$ is largest at $\wt{\Delta t}
= -0.22$.
%
% It is sharpest (largest $\left. \d \wt{\omega_z}(s) / \d
% s \right|_{s=0}$) at $\wt{\Delta t} = -0.22$.
%
The result might illustrate causality: it is the {\em past} velocity field
that has shaped the passive scalar field $\varphi$ and the \tnti\
that is obtained from $\varphi$.
% \marginpar{JW: not sure about this; might be an artefact of the interrogation algorithm that puts the velocity at the location of the first interrogation window, rather than the middle of the two interrogation windows}
%
% That a time shift should indeed
% result in a change of the conditional average was already anticipated
% in Fig.\ \ref{fig.stream0.2}.
%
\nonote{What to make of this time jump.  It is not entrainment
velocity.  Also, it is not in the conditional average of $\Psi_0$...}
\nonote{Ali: {Westerweel2005} is for a different Reynolds number, it is
not averaged over $x$, while its horizontal axis is zoomed out.
Still, it looks prettty similar.}

Conditional averages with respect to the \tti\ are shown in Fig.\
\ref{fig.res.omega}(c);  
%%% The result is surprising: on \tti s $\wt{\omega_z}$ jumps higher, with a more outspoken superlayer bump \citep{Westerweel2009}. 
this result shows a more significant jump of the conditional vorticity across the \tti\ interface, as well as a more pronounced peak that is indicative of a shear layer at the \tti\ \citep{Westerweel2009}. 
This is a rather surprising result, since
%%% However, 
this internal interface is no longer defined
as one between vorticity and its absence, but between two levels of
%%%
conditional vorticity associated with two distinct levels of 
scalar concentration.  
%
% That 
This result of 
an increased turbulence level yielding 
%%% larger structure of
a more pronounced jump in
conditional averages agrees with the findings of \citet{Eisma2015pof}
in a turbulent boundary layer.

%--------------------------------------------------------------------
\begin{figure}[t]
\centering
\includegraphics[scale = 0.9]{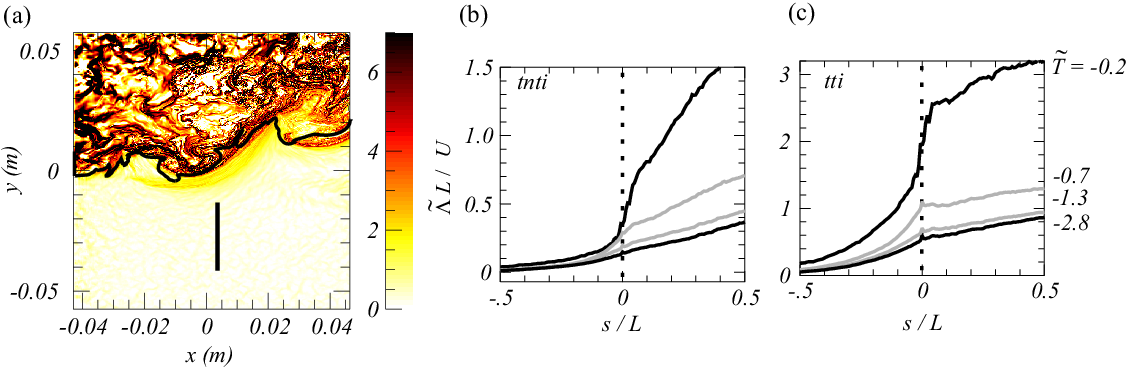}
% see version 1, with time shifts
%
\caption{% Conditional averages of $\Lambda_{-T}$.  
(a) Snapshot of the field $\Lambda_{-T}$ for $T = -$2.8$L/U_c$. The black curve represents the 
\tnti\ as detected from the fluorescent dye.
% (black), \tti\ (cyan).  
The vertical bar indicates the extent of the horizontal axis for $-0.5\leq s/L \leq +0.5$ in
panels (b,c).
%
% (b) Conditional average of $\Lambda_{-T}$.  Indicated is th influence
% of the time delay $\Delta t$ between the scalar field
% $\varphi(\vec{x}, t)$ and $\omega(\vec{x}, t + \Delta t)$, and thus
% between perpendicular intersections of the \tnti\ at $t$, and
% $\Lambda_{-T}$ at $t + \wt{\Delta t}$, with $\Delta t = 0.22\:L/U$.
%
%
(b) Conditional averages of $\Lambda_{-T}$ along normals on the
\tnti\ for a range of (backward) integration times $-T$ expressed in
$L / U_c$ = 0.38~s. 
(c) Same as panel (b), but now for the \tti.  For the curves in
(b,c) the asymptote for $s \ll -L$  (i.e., into the irrotational domain) is
set to 0.
}
\label{fig.res.sig}
\end{figure}
%--------------------------------------------------------------------
% \marginpar{in (b) and (c) change $-.5$ to $-0.5$}

% The superlayer is a thin shear layer at the outer edge of the jet.
% As we will show below, it can also be found at \tti s.

%..............................................................Lambda
\subsection{Conditional averages of $\Lambda_{T}$}
In the following figures we show a snapshot of the 
% \marginpar{what does `same' refer to?}
barriers to the advective  
field (the
first one of the 1.6$\times$10$^3$ frames taken) together with the
conditional averages, both on the \tnti\ and the \tti.  The
quantitive results of the conditional averages are based on all
frames in all repeated runs.  Of course, the frames in a single run
are not 
%%%
statistically 
independent.
The choice for backward times is inspired by the results of
\citet{Jesse2021} who found a relation with the edges of uniform
concentration zones.
\nonote{This all passed our ``random tests'', see ``Extra results'',
which will not go in the paper.}

We show snapshots of the Lyapunov field $\Lambda_{-T}$ and the
diffusive flux field $\Psi_{-T}$ at $T$~=~$-$2.8$L/U_c$. For the
conditional averages, the (backward) integration times varied from 
$T=-$0.2$L/U_c$ to $T=-$2.8$L/U_c$, with the time scale $L / U_c$
= 0.38~s now taken at the start ($x$ = 0.57~m) of a
run.
The actual integration times are limited by the residence time of
fluid parcels in the moving observation frame;
see Fig.\ \ref{fig.setup}(c).

Fig.\ \ref{fig.res.sig}(a) shows 
%%%
an example of 
the Lyapunov field $\Lambda_{-T}$.
%%% repeat 
% The integration time $T$ for this snapshot depends on the location in the frame, as shown in figure~\ref{fig.setup}(d). 
%%%
Since the integration time $T$ depends on the location in the frame,
%%% Clearly, 
longer
integration times result in sharper features 
%%% if 
when barriers 
%%% stay
remain
invariant in time. Since the traversing velocity of the cameras is 
%%% optimized for 
set to follow 
the \tnti,
%%% so that 
the features of $\Lambda_{-T}$ (and $\Psi_{-T}$) 
%%% are 
become 
increasingly
blurred towards the core of the jet.

The conditional averages of $\Lambda_{-T}$ 
%%% with respect to 
along the local normal directions of 
the \tnti\
%%% on interface-normals is 
are 
shown in Fig.\ \ref{fig.res.sig}(b). The
dependence on the normal distance $s$ to the \tnti\ suggests a
correlation between the finite-time Lyapunov field and the \tnti,
especially for the shorter integration times.  The correlation decreases
with increasing $T$, and appears to reach an asymptote at the longest
integration time.  At that time, the correlation is weaker than the
correlation with the vorticity field. 
%
% There no longer is a marked dependence on the delay time $\Delta t$.
% This is not surprising as $\Lambda_{-T}$ involves a long-time
% average.
%
% \note{Could scale the height of the peaks, but what does it mean ?}
These trends 
%%% are 
appear to be 
much stronger for 
%%%
conditional 
averages with respect to the \tti,
which are shown in Fig.\ \ref{fig.res.sig}(c).
%
% \note{Asaad, Jerke: stronger interface with more turbulence?}

%--------------------------------------------------------------------
\begin{figure}[t]
\centering
\includegraphics[scale = 0.87]{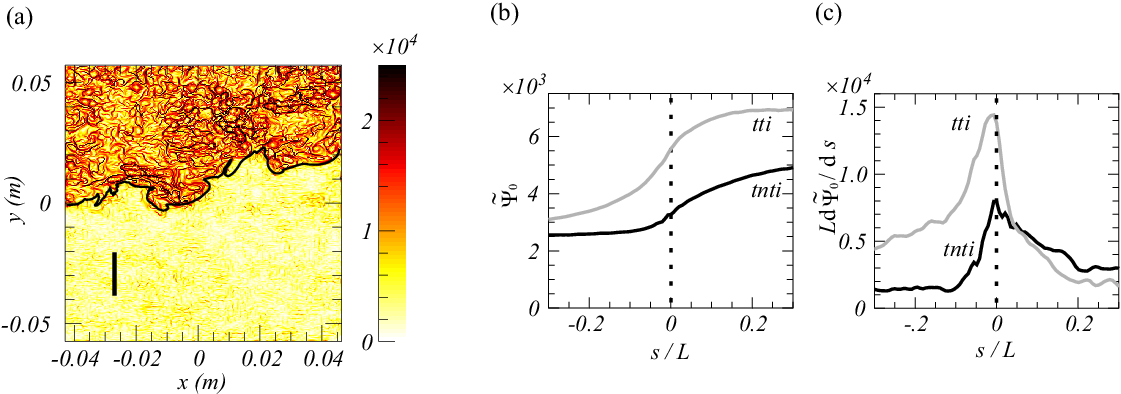}
\caption{Conditional averages of $\Psi_{0}$.  
(a) Snapshot of the zero-time diffusive barrier field $\Psi_0$; the black line represents the
\tnti\ detected from the fluorescent dye. The vertical bar indicates the extent of the horizontal axis for $-0.3\leq s/L \leq +0.3$
in panels (b) and (c).
(b) Conditional average of $\Psi_{0}$ over perpendicular
intersections of the \tnti\ and \tti. (c) Slopes of the curves in panel (b) sharply peak at the interface locations. 
%
% (c) Same as panel (b), but with a different vertical scale. Gray
% line: vertical conditional average. Thin black lines: time delay
% $\wt{\Delta t} = \pm 0.22$.
% 
% Indicated is the influence of the time delay $\Delta t$ between the
% scalar field $\varphi(\vec{x}, t)$ and $\omega(\vec{x}, t + \Delta
% t)$, and thus between perpendicular intersections of the \tnti\ at
% $t$, and $\Lambda_{-T}$ at $t + \wt{\Delta t} = 0.22\:L/U$. 
%
}
\label{fig.res.psi0}
\end{figure}
%%% \marginpar{show derivative of fig. 8(b), with original data in inset; do the same for Fig. 7(b-c) and Fig. 9(b-c) ??}
%--------------------------------------------------------------------

%.................................................................Psi
\subsection{Conditional averages of $\Psi_{0}$ and $\Psi_{-T}$}
\label{sec.res.psi}
Before discussing the diffusive barrier fields $\Psi_{-T}$ and
$\Psi_{0}$, we detail some technicalities.  Visualization of the
associated structures requires the integration of a dynamical system.
For the equal-time diffusive barrier field $\Psi_0(\vec{x}_0, t_0)$
%%% it is
we have
\be
   \frac{\d \vec{x}}{\d \xi} = \vec{h}(\vec{x}(\xi); t_0), \quad
   \text{with:}\quad
   \vec{h} = \nu \nabla^2 \vec{u}, 
   \quad\text{and:}\quad
   \vec{x}(\xi = 0;t_0) = \vec{x}_0,
\label{eq.s}
\ee
where the active variable (dimensionless `pseudo time') $\xi$ is a
curvilinear coordinate, and $t_0$ is a parameter 
%%% : 
that represents
the physical time
at which the structure of $\vec{h}(\vec{x_0}, t_0)$ is computed.  The
evolution of the vector field $\vec{x}(\xi)$ over a (pseudo) time
interval $\Xi$ can be described by a flow map ${\cal F}$:
$\vec{x}(\Xi) = {\cal F}_0^\Xi (\vec{x}(0))$.  Much as in the case of
the 
%%% FTLE, 
finite-time Lyapunov exponent, 
its gradient $\mat{M}_0^\Xi = \nabla {\cal F}_0^\Xi$
defines a Cauchy-Green tensor $\mat{C}_0^\Xi = \mat{M}_0^\Xi
\left(\mat{M}_0^\Xi \right)^T$.  The diffusive barrier field $\Psi_0$
is the logarithm of its largest eigenvalue.  
The numerical integration of Eq.~(\ref{eq.s}) is done over $\Xi$ =
2.5$\times$10$^{-4}$, corresponding to a displacement $\Delta x$ = 
1.4$\times$10$^{-4}$~m, where $\nabla \vec{h}$ is computed from
finite differences.

From the appearance of Fig.\ \ref{fig.psi0.2}, which displays
interesting small scale structures, the effort of integrating a dynamical system
for visualization, which may look cumbersome at first sight, is
worthwhile.

\nonote{what about the sign of $\Psi$ ?  What about its units ?}

%--------------------------------------------------------------------
\begin{figure}[t]
\centering
\includegraphics[scale = 0.9]{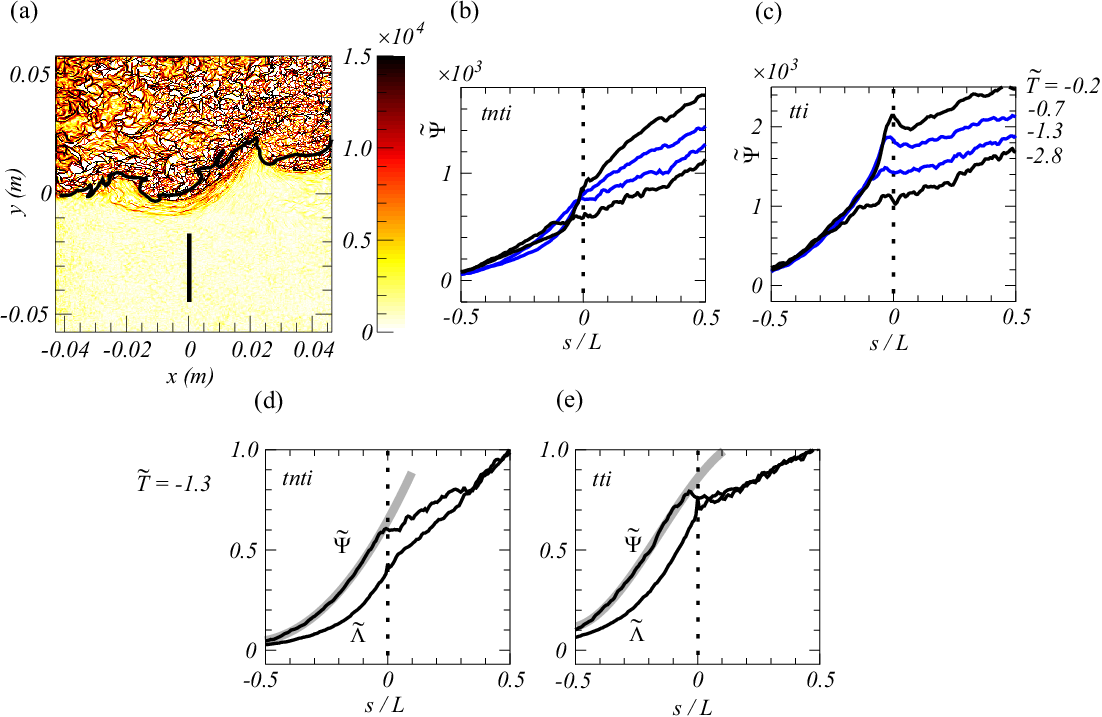}
\caption{% Conditional averages of $\Psi_{-T}$.  
(a) Diffusive barrier field $\Psi_{-T}$. The black curve represents the \tnti\ detected through the fluorescent dye.  The vertical
bar indicates the extent of the horizontal axis for $-0.5\leq s/L \leq +0.5$ in panels (b-e).
(b) Conditional average of $\Psi_{-T}$ over normal sections of
the \tnti\ for a range of (backward) integration times $-T$, 
expressed in $L / U$ = 0.38~s.  The curves asymptote to a
background value for $s \ll -L$; this value was subtracted.  
%
% Indicated is the influence of the time delay $\Delta t$ between the
% scalar field $\varphi(\vec{x}, t)$ and $\omega(\vec{x}, t + \Delta
% t)$, and thus between perpendicular intersections of the \tnti\ at
% $t$, and $\Lambda_{-T}$, with ${\Delta t} = -0.22, 0$ and
% $0.22\:L/U$. 
%
(c) Same as (b), but for the \tti.  
(d) Normalized conditional averages $\Psi_{-T}$ and $\Lambda_{-T}$ on
the \tnti\ at $\wt{T} = 1.3$.  The normalization is such that the
asymptote at small $s$ is set to 0 and the value at $\wt{s} =
0.5$ is set to 1. 
(e) Same as (d), but for the \tti.
The gray lines are polynomial fits to guide the eye that become linearly dependent on $s$ for $s \ge 0$.
}  
\label{fig.res.psi}
\end{figure}
%--------------------------------------------------------------------

The zero-time field $\Psi_0$ and its conditional average are shown in
Fig.\ \ref{fig.res.psi0}, with a 
%%% blowup 
larger view 
already shown in Fig.\ \ref{fig.psi0.2}.
% 
% The average over perpendicular cross sections in Fig.\
% \ref{fig.psi0.2}(c) differs from that of the vertical sections in a
% way similar to that of the vorticity in Fig.\ \ref{fig.res.omega}(b).
% 
Since $\Psi_0$ involves 
%%%
the computation of 
a second derivative, the noise in the
irrotational domain is now larger than that of $\wt{\omega_z}$.
% The influence of the time shift between $\vec{x}_{\rm tnti}$ and
% $\Psi_0$ is much smaller than for the vorticity field.
%
Despite 
%%% its 
this
elevated noise level, the field $\Psi_0$ sharply defines
the turbulent domain indicating that Lyapunov operator amplifies the signal to noise ratio. Clearly, the representation $\Psi$, which
entails the curvature properties 
% \marginpar{is that relation already mentioned in Sect. II?}
of the the streamlines of the vector
field $\overline{\vec{b}}_{t_0}^t$, has a regularizing effect.

Compared to the conditional vorticity, the conditional average $\wt{\Psi}_0$ lacks the impression of a `superlayer'
\citep{Westerweel2005}.  Much as for the vorticity, conditional
averages with respect to the \tti\ show a larger 
%%%
jump across the interface.

The finite-time diffusive flux field $\Psi_{-T}$ is shown in Fig.\ \ref{fig.res.psi}(a).
Compared to the zero-time field $\Psi_0$ it
involves the time-averaged vector field 
$\overline{\vec{b}}_{t_0}^{t_0 + T}$, which is visualized in the
same way as the field $\vec{h}(\vec{x}, t) = \nu \nabla^2
\vec{u}(\vec{x}, t)$ in the case of $\Psi_0$.  The conditional
average in Fig.\ \ref{fig.res.psi}(b) evolves into a featureless
asymptote for increasing integration times $T$.  
%
% The effect of the second derivative of the measured velocity field
% leads to an elevated noise level, which was subtracted.  
%
In comparison with the conditional average of the advective barrier
field $\Lambda_{-T}$ field in Fig.\ \ref{fig.res.sig},
$\wt{\Psi}_{-T}$ reaches its asymptote at shorter times $T$.
The faster decorrelation of diffusive structures is primarily because diffusion occurs on a smaller time scale compared to advective structures.

Correlation of the fields $\Lambda$ and $\Psi$ with interfaces would
show up as (sharp) jumps in their conditional averages at $s = 0$.  To
highlight the differences and similarities of $\wt{\Lambda}_T$ and
$\wt{\Psi}_T$ at $\wt{T}$ = -1.3 (the tilde symbol refers to a scaled form), 
% \marginpar{why does $T$ suddenly has a tilde? $T$ was already defines wrt $L/U_c$.}
they are shown normalized in 
figure~\ref{fig.res.psi}(d,e); their asymptotes at small $s$ is set to 0
while the value at $\wt{s}$ = 0.5 is set to 1. The correlation of
both fields with the \tti\ is slightly stronger than that for the
\tnti, but otherwise scaled $\wt{\Lambda}_T$ and scaled $\wt{\Psi}_T$ are not
significantly different.
%
% We finally notice that both fields $\Lambda$ and $\Psi$ display large spatial fluctuations. Inside the turbulent domain their conditional average is an order of magnitude smaller than their typical magnitude mainly because of the nature of these structures. Diffusive barriers have bounded shapes while advective barriers are more aligned with the main advection direction (i.e, streamwise direction). Both structures have scalar field where if no barrier is present, the scalar value would be close to zero. This drastically reduces the conditional average values with more severe effect on diffusive barrier conditional averages. 

We observe that both fields, $\Lambda$ and $\Psi$, exhibit large spatial fluctuations. Within the turbulent domain, their conditional averages are an order of magnitude smaller than their typical magnitudes, primarily due to the nature of these structures. Diffusive barriers have bounded shapes, while advective barriers form elongated structures that align with the main advection direction (i.e., the streamwise direction). Both structures are represented by a scalar field, where the scalar value 
%%% closes 
remain close 
to zero in regions without any barriers, even within the turbulent domain. This significantly reduces the conditional average values, with a greater effect on the conditional averages of the diffusive barriers.
%--------------------------------------------------------------------
%\marginpar{fig. (a) should indicate the irrotational and turbulent sides of the interface}
\begin{figure}[t]
\centering
\includegraphics[scale = 0.9]{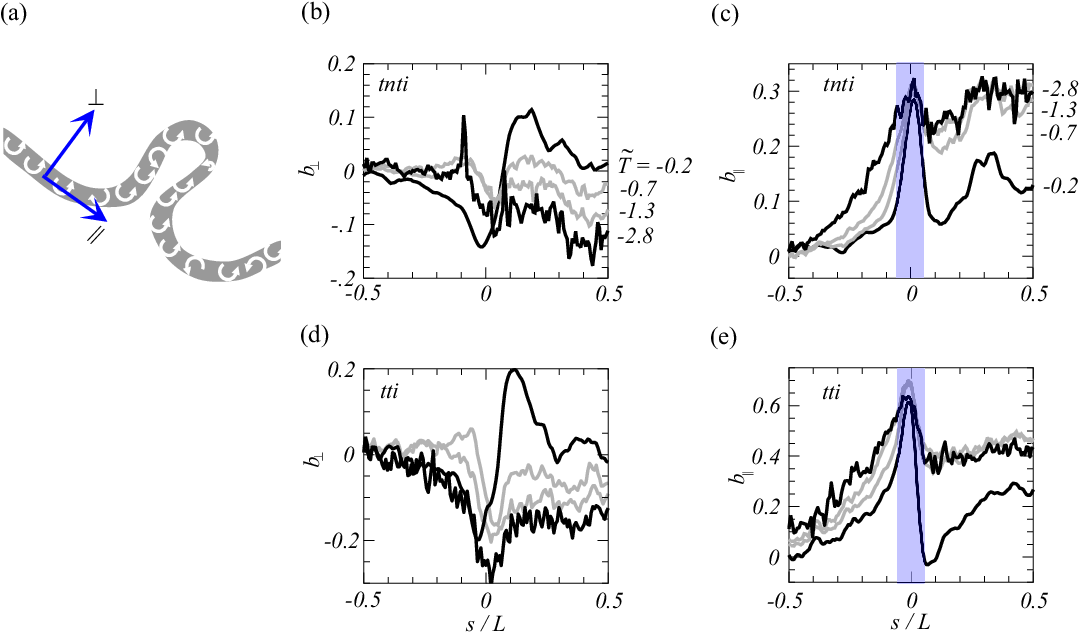}
\caption{% Didffusive flux$.  
(a) Cartoon illustrating the diffusive flux normal ($\perp$) and
tangential ($\parallel$) to an interface.
(b) Conditional average of diffusive flux $b_\perp =
\overline{\vec{b}}_{t_0}^{t_0-T}\cdot \vec{e}_{\rm tnti}$ normal to
the \tnti\ interface, for integration times $\wt{T}$ = $-0.2,\ldots,
-2.8$.
(c) Conditional average of the flux
$b_\parallel = \overline{\vec{b}}_{t_0}^{t_0-T}\cdot \vec{t}_{\rm tnti}$ {\em
parallel} to the \tnti\ interface. 
The width of the blue vertical box indicates the Taylor microscale
$\wt{\lambda}$.
(d, e) Same as (b, c), but for the \tti.
}
\label{fig.flux.psi}
\end{figure}
%--------------------------------------------------------------------

% \marginpar{why is there a tilde for $T$? This seems inconsistent.}
\subsection{Diffusive momentum flux}
While $\Psi_{-T}$ gauges the convergence properties of the averaged
vector field $\overline{\vec{b}}_{t_0}^{t_0-T}$, such that it is
large on lines to which $\overline{\vec{b}}_{t_0}^{t_0-T}$ is
tangent, the diffusive flux through the interface can also be
measured directly.  Conditional averages of the normal flux $b_\perp
= \overline{\vec{b}}_{t_0}^{t_0-T} \cdot \vec{e}_{\rm tnti}$, and its
tangential component $b_\parallel = \overline{\vec{b}}_{t_0}^{t_0-T} \cdot
\vec{t}_{\rm tnti}$, with $\vec{t} \perp \vec{e}$, are shown in Fig.\ \ref{fig.flux.psi}(b,c), respectively.

A striking observation is that the tangential component of the
diffusive flux $b_\parallel$ is concentrated in the diffusive
superlayer.  The width of this superlayer is comparable to the Taylor
microscale; see Sec.~\ref{sec:setup}. The tangential flux remains comparably invariant in time, with the momentum
gradient alight with the flow direction.
% \marginpar{Make sure you give a numerical value in Sect. III}
% That $\overline{\vec{b}}_{t_0}^{t_0-T}$ flows into the irrotational
% domain demonstrates that the \tnti\ propagates outward through
% small-scale nibbling, and is {\em not} a tangency line of
% $\overline{\vec{b}}_{t_0}^{t_0-T}$.
From the finite-time normal diffusive flux through the interface, we observed negative flux upon entering the interface, suggesting that viscous diffusion transports momentum in a way that the interface grows and propagates into the irrotational domain, and is {\em not} a tangency line of $\overline{\vec{b}}_{t_0}^{t_0-T}$. However, unlike the tangential flux, the normal flux is not invariant in time; it increases as the integration time of the Lagrangian diffusive flux increases.
These properties are more outspoken for the \tti, as shown in 
figure~\ref{fig.flux.psi}(d,e).
%%
%%
% A cartoon illustrating these fluxes is shown in
% Fig.\ref{fig.flux.psi}(a).
%
The fluctuation amplitudes of these fluxes increase with increasing
integration time.  This may explain the diminishing correlation of
$\Psi_{-T}$ with the interfaces. 
% \marginpar{I miss the discussion that the decorrelation is due to out-of-plane motion while all integrations are in the same measurement plane.}

%....................................................................
\section{Conclusions}
\label{sec:conclusions}
%
% We define the \tnti\ as the border between dye and its absence in the
% seeded turbulent efflux of the jet.  We proposed a refinement of the
% edge detection based on cluster analysis of the measured scalar
% concentration field.  We emphasize the importance of perpendicular
% intersections of the interface in conditional averages.
%
We investigate the transport mechanisms of turbulent advection and
viscous diffusion, as shaped by turbulent interfacial layers, with
the outermost one the \tnti.  High spatial resolution and long
observation times are achieved in an experimental setup where the
field of view moves with the interface, providing quasi-Lagrangian
information.

% Our experiment, where we move with the \tnti, gives access to coherent
% structures that exist over long times.  Of those, the FTLE field
% that is associated with large-scale motion still correlates with 
% turbulent interfaces, but not the field $\Psi_T(\vec{x})$, which
% expresses blockage of diffusive momentum transport.
% 
% This is in accordance with the interpretation of ridges of the FTLE
% field that block large-scale momentum transport, and the diffusive
% barrier field whose maxima block diffusive momentum flux.  It agrees
% with the idea of ``nibbling'' where the turbulent domain grows
% through diffusive transport of vorticity.

%
Objectively, we have identified edges in a turbulent velocity field
from the distribution of an advected passive scalar.  These edges, be it the
interface between turbulence and the surrounding quiescent fluid or an internal edge in
the turbulent domain, act as shear layers with an associated
concentration of vorticity. The surprise is that this works with a
rather arbitrary choice of the number of concentration clusters (here $n_c$
= 4). Other works \citep{Eisma2015pof, Holzner2006, Asadi2022} also
find these edges, but using contours of vorticity or enstrophy
instead of scalar concentration. 

Our findings show that the intensity of Lagrangian advective and
%%% of 
diffusive terms correlate with the interfacial layers. However,
these correlations diminish as the integration time $T$ increases, with
barriers to viscous flux decorrelating faster than advective
barriers. Most probably, this difference arises because viscous diffusion is a
small-scale process, whereas advection occurs on larger scales.
Therefore, these Lagrangian structures should decorrelate over
shorter times as their scale decreases. This is in accordance with
the interpretation of ridges of the finite-time Lyapunov exponent (FTLE) field that block large-scale
momentum transport, and the diffusive barrier field whose maxima
block diffusive momentum flux.

Our experiment, where we move with the flow, allows us to study the influence of the observation time $T$. Remarkably, averaging over longer times $T$ results in noisier curves (see Fig.\ \ref{fig.flux.psi}). Perhaps longer Lagrangian trajectories encounter more large gradients.  

A caveat is that our analysis is two-dimensional as it is based on 2D planar measurement of the velocity and scalar fields. Over one large-eddy turnover time, Lagrangian tracks may wander away from the measured plane, leading to inevitable decorrelation.

The diffusive flux at the interface agrees with the idea of so-called `nibbling' where the turbulent domain grows outward through viscous diffusion transport of vorticity and is illustrated vividly in Fig.\ \ref{fig.flux.psi} persistent flux as time progresses. 
%
%A diffusive superlayer, with a width comparable to the Taylor microscale and relatively invariant over time, appears to concentrate the diffusive flux parallel to the \tnti.
The diffusive flux parallel to the \tnti\ is localized within a superlayer whose width is comparable to the Taylor microscale and remains relatively invariant over time.
%

% The idea of `nibbling,' in which the turbulent domain grows outward through the viscous diffusion transport of vorticity, is consistent with the observed diffusive normal flux at the interface and is illustrated vividly in figure~\ref{fig.flux.psi} persistent flux as time progresses. 

% \marginpar{This should be quantified!!}

% \nonote{The Taylor scale discussion is moved up}

% Surprisingly, the same conclusions hold for the \tti\ that was
% identified in our experiments.  Therefore, a turbulent interface
% between rotational and irrotational fluid is equivalent to one
% between two distict levels of turbulence.  Conversely, the
% conditional averages in Figs.\ \ref{fig.res.omega},\ref{fig.res.sig}
% and \ref{fig.res.psi}, demonstrate the relevance of uniform
% concentration zones, as it is these zones that define our \tti.

%--------------------------------------------------------------------
\begin{acknowledgments}
We acknowledge the support and expertise of Ing. Edwin Overmars in performing the PIV and LIF measurements. 
\end{acknowledgments}

%....................................................................
\setlength{\bibsep}{1ex}
\setlength{\bibhang}{2ex}
% \bibpunct{(}{)}{;}{a}{ }{,} % uncomment: author-year
% = style of referencing, use (), use ; in multiple cites, a: use
% or use \setcitestyle{authoryear,round,sort}
% author-year, use space between author names and year
% anyway, prf wants numbered

\bibliography{Transport}
%--------------------------------------------------------------------
%--------------------------------------------------------------------

\end{document}